\documentclass[aip,jcp,notitlepage,twocolumn,reprint,floatfix]{revtex4-2}

\usepackage{graphicx}
\usepackage{dcolumn}
\usepackage{bm}
\usepackage{hyperref}
\hypersetup{colorlinks=true, citecolor=blue, urlcolor=blue, linkcolor=blue}
\usepackage{amsmath}
\usepackage{mathtools}
\usepackage{booktabs}
\usepackage{multirow}
\usepackage{comment}
\usepackage{natbib}
\usepackage{xcolor}
\usepackage{soul}
\let\oldtheequation\theequation
\makeatletter
\def\tagform@#1{\maketag@@@{\ignorespaces#1\unskip\@@italiccorr}}
\renewcommand{\theequation}{(\oldtheequation)}
\makeatother

\newcommand{\appref}[1]{\hyperref[#1]{Appendix~\ref{#1}}}

\begin{document}

\title{Multireference protonation energetics of a dimeric model of nitrogenase iron--sulfur clusters}

\author{Huanchen Zhai}
\email{hczhai.ok@gmail.com}
\affiliation{Division of Chemistry and Chemical Engineering, California Institute of Technology, Pasadena, CA 91125, USA}

\author{Seunghoon Lee}
\affiliation{Division of Chemistry and Chemical Engineering, California Institute of Technology, Pasadena, CA 91125, USA}

\author{Zhi-Hao Cui}
\affiliation{Division of Chemistry and Chemical Engineering, California Institute of Technology, Pasadena, CA 91125, USA}

\author{Lili Cao}
\affiliation{Department of Theoretical Chemistry, Lund University, P. O. Box 124, SE-221 00 Lund, Sweden}

\author{Ulf Ryde}
\email{ulf.ryde@compchem.lu.se}
\affiliation{Department of Theoretical Chemistry, Lund University, P. O. Box 124, SE-221 00 Lund, Sweden}

\author{Garnet Kin-Lic Chan}
\email{gkc1000@gmail.com}
\affiliation{Division of Chemistry and Chemical Engineering, California Institute of Technology, Pasadena, CA 91125, USA}

\begin{abstract}

Characterizing the electronic structure of the iron--sulfur clusters in nitrogenase is necessary to understand their role in the nitrogen fixation process. One challenging task is to determine the protonation state of the intermediates in the nitrogen fixing cycle. Here, we use a dimeric iron--sulfur model to study relative energies of protonation at C, S or Fe. Using a composite method based on coupled cluster and density matrix renormalization group energetics, we converge the relative energies of four protonated configurations with respect to basis set and correlation level. We find that accurate relative energies require large basis sets, as well as a proper treatment of multireference and relativistic effects.
We have also tested ten density functional approximations for these systems. Most of them give large errors in the relative energies. The best performing functional in this system is B3LYP, which gives mean absolute and maximum deviations of only 10 and 13 kJ/mol with respect to our correlated wavefunction estimates, respectively, comparable to the uncertainty in our correlated estimates.
Our work provides benchmark results for the calibration of new approximate electronic structure methods and density functionals for these problems.
\end{abstract}

\maketitle

\section{Introduction}

Nitrogenase is the only enzyme that can catalyze the conversion of atmospheric dinitrogen (\( \mathrm{N_2} \)) to ammonia (\( \mathrm{NH_3} \)), the key reaction in nitrogen fixation.\cite{beinert1997iron,tanifuji2020metal,seefeldt2020reduction,rutledge2020electron} Extensive biochemical research has revealed that the catalysis in Mo-nitrogenase takes place in the MoFe-protein, which contains a FeMo-cofactor (FeMoco) cluster, with composition \( \mathrm{MoFe_{7}S_{9}C(homocitrate)} \), responsible for the \( \mathrm{N_2} \) reduction, and a P-cluster, with  composition \( \mathrm{[Fe_8S_7Cys_6]} \), that transfers electrons to the FeMoco active site.\cite{kirn1992crystallographic,peters1997redox} During the last two decades, the atomic structures of the nitrogenase clusters have been determined by X-ray crystallography.\cite{einsle2002nitrogenase,spatzal2011evidence,lancaster2011x} Given these structures, \emph{ab initio} electronic structure computation may be applied to determine the binding sites, reaction intermediates and eventually the catalytic mechanism.\cite{jafari2022benchmark,benediktsson2022analysis}

Recently, the intriguing \( \mathrm{E_4} \) intermediate of nitrogenase,\cite{hoeke2019high} formed by adding four electrons and protons to the  \( \mathrm{E_0} \) resting state of FeMoco and responsible for the binding of N$_2$, has been studied computationally by several groups.\cite{Hoffman:18,Einsle:14,Siegbahn:16,Dance:20,cao2018protonation,cao2020structure,thorhallsson2019model} This is a formidable task owing to the numerous possible binding positions of the added protons\cite{cao2018protonation} and the complicated electronic structure of the FeMo cluster.\cite{Noodleman:21,Cao:18-bs}  All the above calculations have been performed using density functional theory (DFT).\cite{hohenberg1964inhomogeneous} However, due to the open-shell and multireference nature of the nitrogenase clusters, the reliability of the obtained DFT results has been called into question and the various functionals predict remarkably different results (over 600 kJ/mol difference in the predicted stability of different protonation states for the E$_4$ state).\cite{cao2019extremely,wei2021active,benediktsson2022analysis} 
In spite of the large number of open-shell transition metal centers in these clusters, it has been shown that approximate full configuration interaction (FCI) methods, such as the \emph{ab initio} density matrix renormalization group (DMRG) algorithm,\cite{white1992density,white1993density,chan2002highly,chan2011density,mitrushenkov2001quantum,chan2016matrix,sharma2012spin,olivares2015ab,wouters2014density,keller2015efficient,baiardi2020density,brabec2021massively,zhai2021low} can tackle the qualitative multireference behavior. For example, earlier calculations using DMRG provided new insights into the electronic structure of the P-cluster with its manifold of low-energy electronic states and their non-classical spin correlations.\cite{sharma2014low,li2019electronic,li2019the}
These studies focused on active space models of the cluster, which were sufficient for a qualitative understanding of the electronic landscape.
However, correctly modeling protonation energetics in FeMoco requires calculations that go beyond qualitative accuracy. In particular, quantitative energetics requires a treatment of electron correlation beyond the strongly correlated active space. 

In this work, we study the protonation problem in the simpler case of a dimeric iron--sulfur cluster. We compare four representative protonation sites (C, S, Fe or bridging two Fe ions), and try to estimate the lowest energy site and the energetic ordering. We do this within a composite approach where we separately treat multireference electron correlation using DMRG and dynamic correlation beyond the active space using coupled cluster methods.
For comparison, we also include several density functional approaches.
Overall, we find that multireference effects and correlation in large basis sets are both crucial to describing the protonation energetics, with correlation effects beyond (perturbative) triples being large. Capturing both effects accurately remains challenging within the composite treatment, but we reach sufficient precision to identify the lowest energy protonation site, as well as to order the relative energies of the protonated structures. In contrast, most of the density functionals that we examine yield qualitative and large quantitative errors for this problem.

\section{Methods}

We consider the dimer  \( [\mathrm{(SCH_3)_2FeS(CH_2)Fe(SCH_3)_2}]^{4-} \)  as a simple model for studying protonation energetics in an iron--sulfur cluster. For this purpose, we added a proton in the following locations: (a) on the bridging CH$_2^{2-}$ group, (b) on the bridging S$^{2-}$ ion, (c) terminally on one Fe ion, or (d) bridging both Fe ions. These locations are representative of potential protonation sites in FeMoco.\cite{cao2018protonation}  
We denote the four protonated configurations  HC, HS, HFe and \( \mathrm{HFe_2} \), respectively. Our goal is to identify the lowest-energy structure and to predict the relative energetics of the four structures.

\noindent \textbf{Structures}. We optimized structures for the four protonated configurations using a broken-symmetry (BS) open-shell singlet ground-state with two antiferromagnetically coupled high-spin Fe(II) centers (net charge --3). Geometry optimization was carried out at the DFT level with the TPSS functional,\cite{tao2003climbing} the def2-SV(P) basis,\cite{schafer1992fully} and using DFT-D3\cite{grimme2010consistent} as a dispersion correction. We show the optimized geometries in \autoref{fig:dimer} and give the coordinates in Supplementary Material Section I. In the optimized structures, we evaluated \(2\langle S_z\rangle \) at each Fe center; these had opposite signs on the different centers and a magnitude ranging from 2.7 to 3.4 depending on the structure. The structures were first designed and used in this work.

\begin{figure}[!htbp]
  \includegraphics[width=\columnwidth]{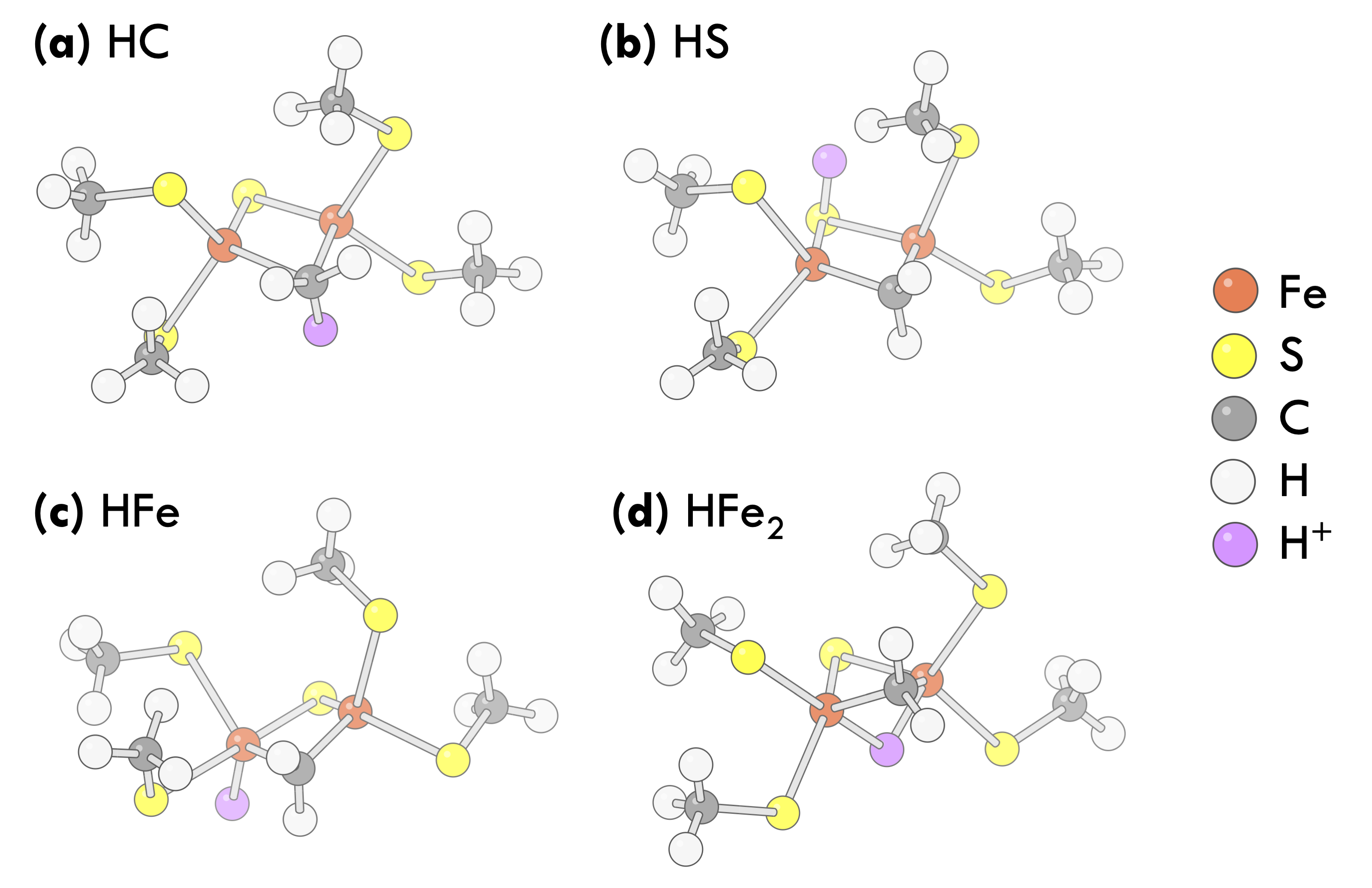}
  \caption{The geometries of the protonated  dimer complex \( \mathrm{[HFe_2S(CH_2)(SCH_3)_4]^{3-}} \) with the added proton on (a) C, (b) S, (c) terminally on one Fe atom and (d) bridging both Fe atoms. The added proton is shown in purple. }
  \label{fig:dimer}
\end{figure}

We summarize the Fe--ligand bond lengths in the four structures in \autoref{tab:geom}. We see that the Fe--ligand distances vary substantially depending on the protonation site. In particular, protonation of the bridging S$^{2-}$ or CH$_2^{2-}$ ions (thereby decreasing their charge to --1) increases their bond lengths to Fe by $\sim$0.2 \AA. Adding the proton terminally to Fe1 (formally yielding a hydride and Fe(IV)) also increases the bond lengths between this Fe ion and its other ligands. However, when the added proton bridges the two Fe ions, the bond lengths are not much changed.

\begin{table}[!htbp]
    \centering
    \caption{The bond lengths (in \AA) between the iron ions (Fe1 and Fe2) and their direct ligands in the optimized geometries of the four dimer models. S$_{\rm t1}$--S$_{\rm t4}$ are the sulfur atoms of four terminal SCH$_3^{-}$ groups, S$_{\rm br}$ is the bridging S$^{2-}$ ion, C$_{\rm br}$ is the carbon atom of the bridging CH$_2^{2-}$ ion and H is the added proton, when binding to Fe.  }
    \begin{tabular}{
        >{\centering\arraybackslash}p{3.4cm}
        >{\centering\arraybackslash}p{1.1cm}
        >{\centering\arraybackslash}p{1.1cm}
        >{\centering\arraybackslash}p{1.1cm}
        >{\centering\arraybackslash}p{1.1cm}
    }
    \hline\hline
    \\
    Bond & HC & HS & HFe & \( \mathrm{HFe_2} \) \\
    \\
    \hline \\
    S$_{\rm t1}$\(-\)  Fe1 & 2.45 & 2.43 & 2.62 & 2.43 \\
    S$_{\rm t2}$\(-\)  Fe1 & 2.45 & 2.45 & 2.50 & 2.48 \\
    S$_{\rm t3}$\(-\)  Fe2 & 2.45 & 2.44 & 2.42 & 2.39\\
    S$_{\rm t4}$\(-\)  Fe2 & 2.42 & 2.39 & 2.38 & 2.43 \\
    \\
    S$_{\rm br}$\(-\)  Fe1 & 2.26 & 2.47 & 2.42 & 2.31 \\
    S$_{\rm br}$\(-\)  Fe2 & 2.28 & 2.53 & 2.21 & 2.34 \\
    \\
    Average (all S\(-\)Fe) & 2.38 & 2.45 & 2.42 & 2.40 \\ 
    \\
    C$_{\rm br}$\(-\)Fe1   & 2.20 & 1.98 &	2.02 & 1.97 \\
    C$_{\rm br}$\(-\)Fe2   & 2.19 & 1.99 &	1.95 & 1.99 \\
    \\
    H\(-\)Fe1              &      &      &	1.76 & 1.72 \\
    H\(-\)Fe2              &      &      &	     & 1.76 \\
        
    \hline\hline
    \end{tabular}
    \label{tab:geom}
\end{table}

To check the effect of the level of theory used for geometry optimization, we compared the optimized geometry for each structure using the def2-SV(P) and def2-TZVP basis sets at the DFT level with the TPSS functional plus dispersion correction (without the X2C relativistic correction). The largest geometric difference in the bond lengths listed in \autoref{tab:geom} between the structures optimized with the two bases is 0.04, 0.09, 0.24, and 0.03 \AA\ for the HC, HS, HFe and \( \mathrm{HFe_2} \) structures. The large difference for HFe is because there is a change in the general structure to another local minimum (the hydride ion and one \( \mathrm{SCH_3} \) ligand change places; this local minimum can also be found with the smaller basis set and then the maximum difference in bond distances is only 0.04 \AA). The changes in the energies relative to HC are 16, 17 and 13 kJ/mol for HS, HFe and \( \mathrm{HFe_2} \), respectively (6 kJ/mol for HFe, when compared to the same local minimum). As these differences are much smaller than the energy difference associated with the choice of protonation site (as discussed later), and in the relevant biological context, the geometries will likely change during the course of dynamics, we will not put too much attention on the exact geometry but simply use the def2-SV(P) optimized geometry in this work.

\noindent \textbf{Composite approach}. For these four structures, we employed a composite energy approach using coupled cluster with singles and doubles (CCSD) and with perturbative triples [CCSD(T)] to estimate dynamic correlation~\cite{bartlett2007coupled,shavitt2009many} and DMRG to estimate multireference correlation. Additional corrections were then included for basis-set completeness and relativistic effects. The final composite energy was computed as
\begin{multline}
    E_\textrm{composite} = E_\textrm{CCSD(T)-TZ}\ + 
    (E_\textrm{DMRG\mbox{-}act} - E_\textrm{CCSD(T)\mbox{-}act} ) \\ + 
    \Delta E_\mathrm{CCSD(T)\mbox{-}CBS} 
\label{eq:composite}
\end{multline}
where \( E_\textrm{CCSD(T)-TZ} \) is the CCSD(T) energy obtained with the cc-pVTZ-DK basis set, $(E_\textrm{DMRG\mbox{-}act} - E_\textrm{CCSD(T)\mbox{-}act} )$ is a multireference correction (estimated in an active space, which will be defined in the following), and $\Delta E_\mathrm{CCSD(T)\mbox{-}CBS}$ is a basis-set correction (specified below). 

The mean-field (HF and DFT) and single-reference post-HF calculations were performed in \textsc{PySCF}.\cite{sun2018pyscf,sun2020recent} Spin-adapted DMRG calculations were performed in \textsc{block2}.\cite{zhai2021low}
For each of these terms, in addition to obtaining energies for the composite energy formula, we carried out additional calculations to understand the impact of different approximations and to estimate the reliability of the corrections. 
We describe the different terms and these aspects below.

\noindent \textbf{Coupled cluster calculations}. For the coupled cluster (CC) calculations, we started from BS unrestricted reference determinants. To understand the influence of orbitals and importance of triples, we first carried out calculations using the small cc-pVDZ-DK basis set and the exact two-component (X2C) scalar relativistic Hamiltonian~\cite{saue2011relativistic,peng2012exact,kutzelnigg2005quasirelativistic} (larger basis set CC calculations are discussed in the basis-set correction section below)
and using 40 frozen core orbitals to reduce computational cost.
To examine the impact of the orbital choice, we used both Kohn--Sham DFT (with the TPSS or B3LYP functionals\cite{becke1988density,lee1988development,becke1993new}), as well as Hartree--Fock Slater determinants. In the unrestricted mean-field calculations, we targeted the projected  \( S_z = 0 \) BS state using an initial guess where the spins in the two Fe atoms were coupled antiferromagnetically.\cite{schurkus2020theoretical} The expectation value of the total \( \langle S^2 \rangle \) in the mean-field state ranged from 3.1 to 5.0 for the structures in this work. Starting from these states, we then computed unrestricted CCSD and CCSD(T) energies.\cite{bartlett2007coupled,shavitt2009many} For the DFT reference determinants, we computed CCSD(T) results based on the semi-canonicalized orbitals [which only diagonalize the occupied-occupied and the virtual-virtual blocks in the Fock matrix. This fixes the definition of the (T) correction]. 

In addition to the low-spin BS state, we also computed the mean-field and CC energies of the high spin (HS) state with \( S_z = 4 \). With this, we estimated the energy of the pure-spin (PS) singlet state (\( S = S_z = 0 \)) from the Yamaguchi formula\cite{yamaguchi1986ab,schurkus2020theoretical} \[ J = \frac{E_{\mathrm{HS}} - E_{\mathrm{BS}}}{ \langle S^2_{\mathrm{BS}} \rangle - \langle S^2_{\mathrm{HS}} \rangle } = \frac{E_{\mathrm{PS}} - E_{\mathrm{BS}}}{ \langle S^2_{\mathrm{BS}} \rangle - \langle S^2_{\mathrm{PS}} \rangle }, \] where $J$ is the exchange coupling. 
For simplicity, we used \( \langle S^2_{\mathrm{PS}} \rangle = 0 \) and \( \langle S^2_{\mathrm{BS}} \rangle \) and \( \langle S^2_{\mathrm{HS}} \rangle \) computed at the CCSD level in the above formula, for computing \( J \) from both CCSD and CCSD(T) energies.
The difference between the BS and PS state can be taken as an estimate of the missing multireference correlation energy arising from spin-recoupling of the Fe centers, but does not capture other types of multireference correlation. 

\noindent \textbf{DMRG multireference correction}. To better estimate the multireference correction, we constructed an active space for a DMRG calculation. We started from a set of (restricted) natural orbitals obtained by diagonalizing the spin-averaged one-particle density matrix (1PDM) of the CCSD wave function calculated above using the cc-pVDZ-DK basis, and then selected orbitals with the occupancy furthest from $0$ or $2$ as the active space. 

Using this active space, we carried out complete active space configuration interaction (CASCI) spin-adapted DMRG\cite{sharma2012spin} calculations, computing the PS singlet state (\( S = 0\)) energies. Before performing DMRG, we split-localized the nearly doubly occupied orbitals, nearly empty orbitals, and other orbitals, using the Pipek–Mezey localization algorithm.\cite{pipek1989fast} The orbitals were then reordered using the Fiedler algorithm.\cite{olivares2015ab} The maximum bond dimension in the DMRG calculations was 5000 [SU(2) multiplets]. We used a reverse schedule to generate data for DMRG energy extrapolation, and the DMRG extrapolation error was estimated as one fifth of the energy difference between the extrapolated
DMRG energy and the DMRG energy computed at the largest bond dimension\cite{olivares2015ab} (see Supplementary Material Section II). For analysis, we also extracted the largest configuration state function (CSF) coefficient from the DMRG wave function, using a deterministic depth-first search algorithm.\cite{lee2021externally}

To obtain a multireference energy correction, we also computed the CCSD and CCSD(T) energies in the same active space, using a BS Hartree--Fock reference.
The initial guess for the active space BS UHF density matrix was obtained by projecting the BS UHF density matrix in the full space into the active space. The correction was then computed as $\Delta E_\mathrm{DMRG\mbox{-}act} = E_\mathrm{DMRG\mbox{-}act} - E_\mathrm{CCSD(T)\mbox{-}act}$.

To validate the size of the correction, we considered three different active spaces, to verify that the multireference effects were converged.
The dimer model in the cc-pVDZ-DK basis contains 180 electrons in 321 spatial orbitals. 
We constructed the active spaces from the UHF/CCSD natural orbitals (see Supplementary Material Section III): one with 36 orbitals and 48 electrons (36o, 48e), one with 55 orbitals and 48 electrons (55o, 48e) and one with 63 orbitals and 64 electrons (63o, 64e).
The uncertainty in the multireference contribution to the relative energies was then estimated crudely as one half the amount of the DMRG multireference correction for the energy difference between HC and \( \mathrm{HFe_2} \) (the least and most multireference structures); here, the one half factor is used to be conservative, as it is the largest estimate of the error still compatible with an assumption that the correction improves the result. We further assume that the DMRG multireference correction computed in the cc-pVDZ-DK basis can be used to correct the CCSD(T) relative energies in the complete basis set (CBS) limit, as shown in \autoref{eq:composite}.

\noindent \textbf{Basis-set correction and relativistic contribution}. To estimate the CBS limit, we used energies computed in several bases: the UHF energy in cc-pVDZ/TZ/QZ-DK bases, as well as CCSD and CCSD(T) energies in the cc-pVDZ/TZ-DK bases.
To estimate the error in the CBS extrapolation, we additionally computed second-order M\o ller–Plesset perturbation theory\cite{moller1934note,pople1976theoretical}
(MP2) energies in the cc-pVDZ/TZ/QZ-DK bases.
To independently analyze the size of the relativistic correction we also computed the CCSD(T)/def2-SV(P) energies with and without using the X2C Hamiltonian. 
The CCSD and CCSD(T) calculations were performed with 40 frozen core orbitals (i.e. excluding the $3s$ and $3p$ semicore on Fe).

Using the CC correlation energies in the cc-pVDZ-DK and cc-pVTZ-DK bases, we extrapolated to the CBS limit energies using the two-point formula\cite{neese2011revisiting}
\begin{align}
E_{\mathrm{corr}}^{\infty} = \frac{X^\beta E_{\mathrm{corr}}^{(X)} - Y^\beta E_{\mathrm{corr}}^{(Y)}}{X^\beta - Y^\beta} \notag
\end{align}
where \( X = 2 \,(\mathrm{DZ}), Y = 3 \,(\mathrm{TZ}) \) and taking \( \beta = 2.4 \). For the corresponding mean-field energy at the CBS limit we simply used \( E_{\mathrm{UHF}}^\infty = E_{\mathrm{UHF}}^{\mathrm{QZ}} \).

To estimate the error in the CCSD(T) relative energies in the CBS limit, we performed an independent extrapolation with the MP2 energies, and took half of the difference between the DZ/TZ extrapolation and TZ/QZ extrapolation (i.e. the difference from the average of the two) using the same CBS formula with \( X = 3 \) and \( Y = 4 \)) for the MP2 energies, namely
\begin{multline}
\Big| \Delta E_{\mathrm{CCSD(T)}}^{\mathrm{DZ/TZ}\rightarrow \infty} - \Delta E_{\mathrm{CCSD(T)}}^{\mathrm{exact}}\Big| \\ \approx
\frac{1}{2}\Big( \Delta E_{\mathrm{MP2}}^{\mathrm{TZ/QZ}\rightarrow \infty} - \Delta E_{\mathrm{MP2}}^{\mathrm{DZ/TZ}\rightarrow \infty}
\Big)
\end{multline}
where \( \Delta E_{\mathrm{CCSD(T)}}^{\mathrm{DZ/TZ}\rightarrow \infty} \) is the difference in the CCSD(T) energies between HC and HFe (the structures showing the largest difference in the extrapolated MP2 energies), estimated from the extrapolation based on DZ and TZ bases to the CBS limit.

We briefly note that we did not use multireference dynamic correlation methods (such as DMRG-second-order $N$-electron valence perturbation theory (DMRG-NEVPT2)) in this work because the  different configurations considered in this work have different bonding topologies. This makes it hard to choose a consistent active space that is also small enough to be used with DMRG-NEVPT2.

\noindent \textbf{DFT comparisons}. For comparison, we computed BS-DFT energies (without the spin-state corrections) using the X2C Hamiltonian and the
TPSS,\cite{tao2003climbing}
BLYP,\cite{becke1988density,lee1988development}
PBE,\cite{perdew1996generalized} 
B97-D,\cite{grimme2006semiempirical}
r$^{2}$SCAN,\cite{furness2020accurate}
TPSSh,\cite{staroverov2003comparative}
B3LYP*,\cite{salomon2002assertion}
B3LYP,\cite{becke1988density,lee1988development,becke1993new}
PBE0,\cite{perdew1996rationale}
and M06\cite{zhao2008m06} functionals, with the cc-pVQZ-DK basis set and with dispersion corrections from the DFT-D3 method.\cite{grimme2010consistent} 

\noindent \textbf{Solvation}. To estimate the effect of solvation, we additionally computed single-point DFT energies using the cc-pVDZ-DK basis, the TPSS functional, the DFT-D3 dispersion correction, the X2C correction, and the domain-decomposition COSMO solvation model\cite{cances2013domain} with a dielectric constant \( \epsilon = 4.0 \). We compared the relative energies with and without the solvation model. We find that solvation greatly stabilizes the negative charges in the model, reducing the number of formally unbound occupied spatial orbitals (i.e. with positive eigenvalues) from 24--25 to less than three, for the four structures. Solvation also leads to a modest, but non-uniform change in the relative energies of the structures (with respect to HC), from \( +2.3 \ \mathrm{kJ/mol} \) for \( E_{\mathrm{HS}} - E_{\mathrm{HC}} \) to \( -16.8 \ \mathrm{kJ/mol} \) for \( E_{\mathrm{HFe}} - E_{\mathrm{HC}} \) and \( -27.1 \ \mathrm{kJ/mol} \) for \( E_{\mathrm{HFe_2}} - E_{\mathrm{HC}} \).
Clearly solvation is important for accurate studies of the biological system. However, its effects can generally be decoupled from that of the correlation level, and thus for the current benchmark study we will henceforth ignore the effects of solvation for simplicity.

\section{Results and discussion}

In \autoref{sec:corr-corr}, we first discuss the CC energies with the cc-pVDZ-DK basis. This will allow  us to understand some features of correlation in the system, including the influence of orbital choice and the size of triples correction on the relative protonation energies, setting the stage for understanding the reliability of the composite method. 
In \autoref{sec:exchange} we discuss the contribution associated with correcting the  BS spin states. In \autoref{sec:dmrg} we discuss the detailed multireference corrections entering the composite energy formula from the DMRG and CC calculations.  In \autoref{sec:basis-set} we discuss CC calculations in larger basis sets, the CBS extrapolation entering the composite energy, and the size of relativistic effects.
We report the final composite energies, the prediction of the lowest energy protonation site and the relative ordering, and the comparison with DFT calculations in \autoref{sec:total-ener}.

\subsection{CC energies: importance of higher-order correlations}
\label{sec:corr-corr}

\begin{table}[!htbp]
    \centering
    \caption{Relative single-point energies (in kJ/mol) for the BS state of the four protonated Fe dimer structures computed using different theories with the cc-pVDZ-DK basis and the scalar relativistic X2C Hamiltonian. The energy of the HC structure is used as the reference.}
    \begin{tabular}{
        >{\centering\arraybackslash}p{3.4cm}
        >{\centering\arraybackslash}p{1.1cm}
        >{\centering\arraybackslash}p{1.1cm}
        >{\centering\arraybackslash}p{1.1cm}
        >{\centering\arraybackslash}p{1.1cm}
    }
    \hline\hline
    \multirow{2}{*}{Theory} & 
    \multicolumn{4}{c}{ Energy difference \( E - E_\mathrm{HC} \) } \\
    & HC & HS & HFe & \( \mathrm{HFe_2} \) \\
    \hline \\
    UHF            & 0.0 & 203.8 & 322.5 & 396.3 \\
    \\
    UHF/CCSD       & 0.0 & 143.7 & 245.3 & 253.1 \\
    UKS-TPSS/CCSD  & 0.0 & 154.9 & 278.9 & 305.9 \\
    UKS-B3LYP/CCSD & 0.0 & 149.3 & 269.6 & 286.8 \\
    \\
    UHF/CCSD(T)       & 0.0 & 134.3 & 190.1 & 179.9 \\
    UKS-TPSS/CCSD(T)  & 0.0 & 132.6 & 179.4 & 163.8 \\
    UKS-B3LYP/CCSD(T) & 0.0 & 132.4 & 185.4 & 167.3 \\
    \\
    \hline\hline
    \end{tabular}
    \label{tab:ener-dz}
\end{table}

\begin{figure*}[!htbp]
  \includegraphics[width=320px]{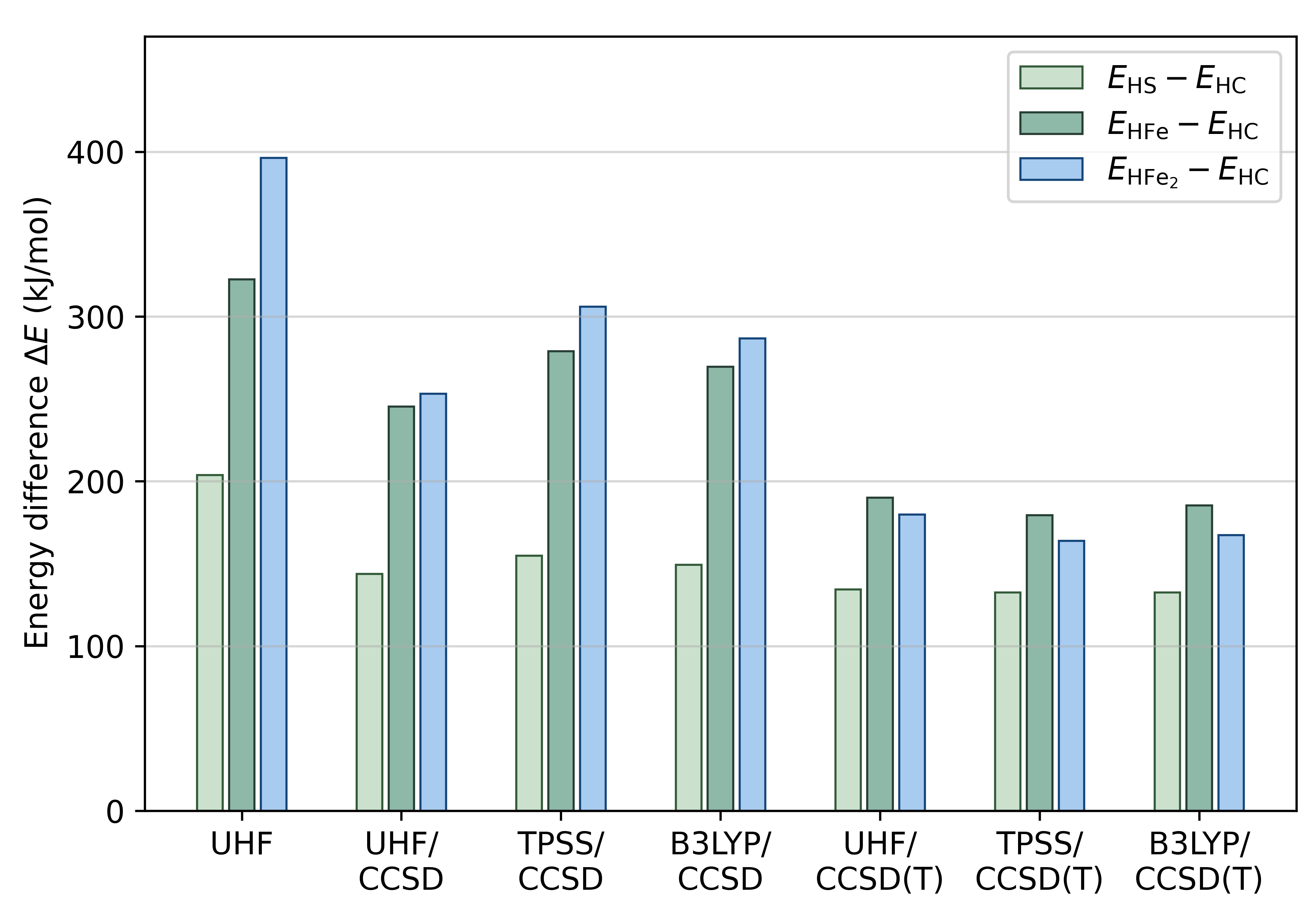}
  \caption{Relative single point energies of the BS state of the protonated Fe dimers computed using different theories with the cc-pVDZ-DK basis and the scalar relativistic X2C Hamiltonian. The energy of the HC struture is used as the reference.}
  \label{fig:ener-dz}
\end{figure*}

We show the energies of the four protonated structures relative to the HC structure from calculations with HF, CCSD and CCSD(T) methods with the cc-pVDZ-DK basis in \autoref{tab:ener-dz} and \autoref{fig:ener-dz}. All methods find that the HC structure is the most stable.
However, CCSD and CCSD(T) predict a different relative ordering.
In addition, there are large quantitative differences in the relative energies, particularly for the HFe and \( \mathrm{HFe_2} \) structures.
For example, the relative energy of the HS structure decreases by 69~kJ/mol on moving from UHF to UHF/CCSD(T), but that of \( \mathrm{HFe_2} \) decreases by 216~kJ/mol.
Comparing the energy differences from UHF/CCSD and UHF/CCSD(T), we see a sizable contribution from (T) to the absolute and relative energies. Specifically, the absolute (T) corrections for the HC, HS, HFe and \( \mathrm{HFe_2} \) structures are \( -214, -224, -269 \) and \( -287 \ \mathrm{kJ/mol} \), meaning that relative to the HC structure, \( \mathrm{HFe_2} \) is further stabilized by triples by as much as 73~kJ/mol (see \autoref{tab:ener-dz}). 
Consistent with this, the CCSD relative energies are observed to be sensitive to the choice of reference orbitals in HFe and \( \mathrm{HFe_2} \).
The large (T) corrections to the relative energies highlight the potentially large contribution of higher-order multireference correlations in the relative protonation energies, especially for the H--Fe bond. 

\subsection{Spin-state corrections}
\label{sec:exchange}

As the above calculations used a BS reference, part of the missing higher-order correlation could potentially originate from the energy difference between the BS and PS singlet states. In \autoref{tab:high-spin} we report the results from the Yamaguchi energy correction to the BS state and the resulting estimate of the PS relative energies. 

The PS state correction to the relative energies is shown in the last two lines in~\autoref{tab:high-spin}. We see that the PS state correction to the relative energies is modest. It is largest for the HFe/HC difference, where it lowers the relative energy by 13 kJ/mol at the CCSD(T) level.
Note that as we explicitly compute multireference contributions from DMRG energies below (which are for PS states), we do not use the PS state energy corrections in the composite energy formula. 

\begin{table}[!htbp]
    \centering
    \caption{Relative single point energies (in kJ/mol) for the BS, high-spin and (estimated) PS singlet states of the protonated Fe dimers computed using different theories with the cc-pVDZ-DK basis and the scalar relativistic X2C Hamiltonian. The energy of the HC structure is used as the reference.}
    \begin{tabular}{
        >{\centering\arraybackslash}p{3.4cm}
        >{\centering\arraybackslash}p{1.1cm}
        >{\centering\arraybackslash}p{1.1cm}
        >{\centering\arraybackslash}p{1.1cm}
        >{\centering\arraybackslash}p{1.1cm}
    }
    \hline\hline
    \multirow{2}{*}{Theory} & 
    \multicolumn{4}{c}{ Energy difference \( E - E_\mathrm{HC} \) } \\
    & HC & HS & HFe & \( \mathrm{HFe_2} \) \\
    \hline \\
    \multicolumn{5}{c}{ High-spin \( S_z = 4 \) } \\ \\
    UHF/CCSD    & 0.0 & 164.1 & 291.6 & 309.0 \\
    UHF/CCSD(T) & 0.0 & 158.5 & 230.7 & 211.9 \\
    CCSD \(\langle S^2\rangle\) & 20.01 & 20.01 & 20.32 & 20.08 \\
    \\
    \multicolumn{5}{c}{ BS singlet } \\ \\
    UHF/CCSD    & 0.0 & 143.7 & 245.3 & 253.1 \\
    UHF/CCSD(T) & 0.0 & 134.3 & 190.1 & 179.9 \\
    CCSD \(\langle S^2\rangle\) & 3.89 & 3.78 & 4.57 & 4.15 \\
    \\
    \multicolumn{5}{c}{ Exchange coupling \( J \) (estimated, \( \mathrm{cm}^{-1} \)) } \\ \\
    UHF/CCSD    & \( -94.0\) & \(-198.2\) & \(-342.0\) & \(-388.5\) \\
    UHF/CCSD(T) & \(-112.2\) & \(-236.2\) & \(-329.9\) & \(-281.5\) \\
    \\
    \multicolumn{5}{c}{ PS singlet (estimated) } \\ \\
    UHF/CCSD    & 0.0 & 139.1 & 231.0 & 238.2 \\
    UHF/CCSD(T) & 0.0 & 128.8 & 177.3 & 171.1 \\
    \\
    \multicolumn{5}{c}{PS singlet correction (estimated)} \\ \\
    UHF/CCSD    & \(-4.4\) & \(-9.0\) & \(-18.7\) & \(-19.3\) \\
    UHF/CCSD(T) & \(-5.2\) & \(-10.7\) & \(-18.0\) & \(-14.0\) \\
    \\
    \hline\hline
    \end{tabular}
    \label{tab:high-spin}
\end{table}

\subsection{Multireference effects}
\label{sec:dmrg}

\begin{figure}[!htbp]
  \includegraphics[width=0.75\columnwidth]{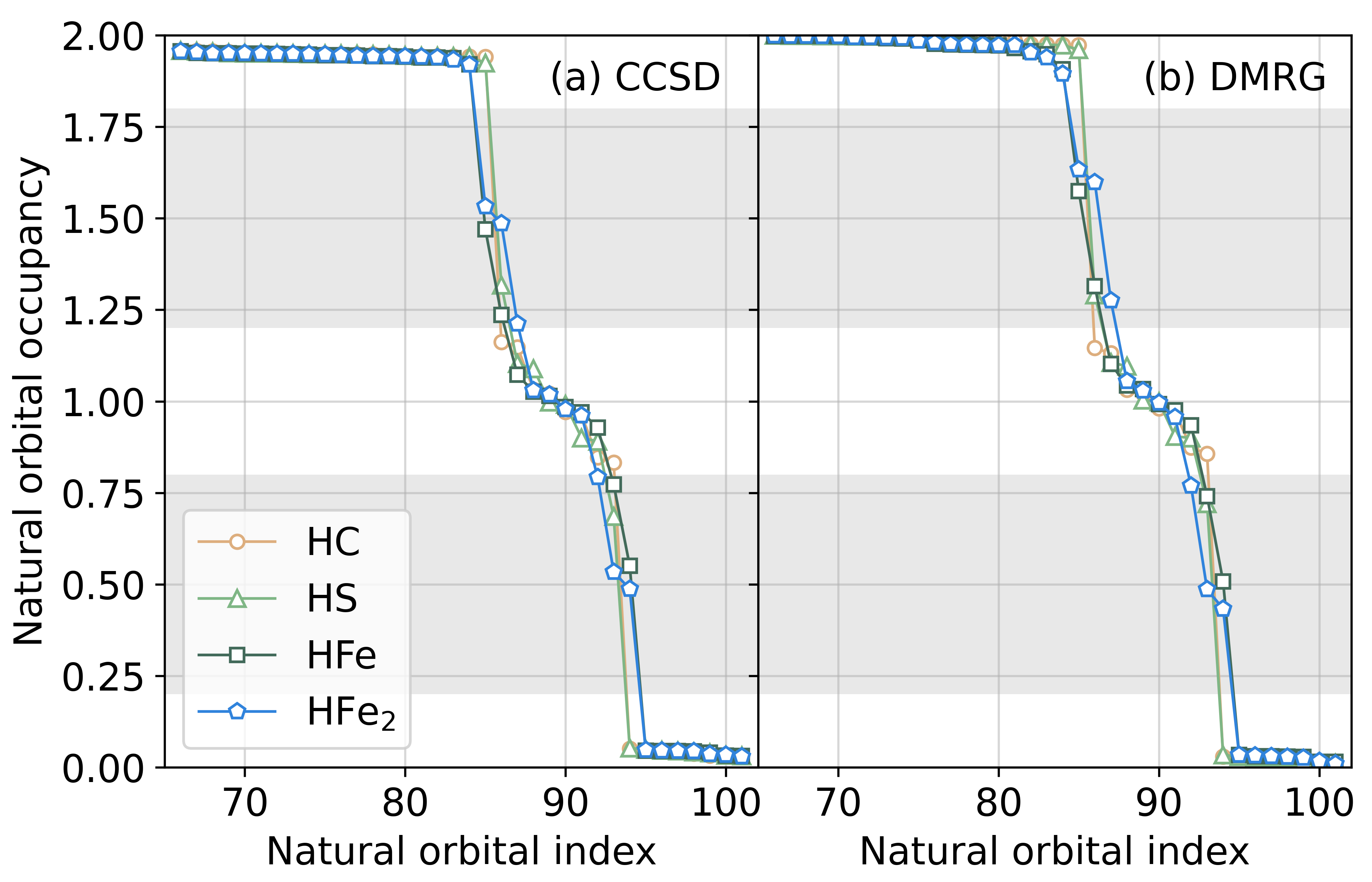}
  \caption{Natural orbital occupancies computed using (a) CCSD, and (b) DMRG in the (36o, 48e) active space, for the four protonated structures. The gray area shows the range of fractional occupancy, as defined in the main text.}
  \label{fig:occ}
\end{figure}

To obtain a more complete picture of the multireference effects, we next consider \emph{ab initio} DMRG energies. 
In \autoref{fig:occ}(a) and (b) we plot the CCSD and DMRG natural-orbital occupancies in the (36o, 48e) active space. We see that the most fractionally and singly occupied orbitals are all included in the active space, which suggests the active space (and its larger counterparts) should capture the  multireference effects in this system. The occupancy patterns of CCSD and DMRG are qualitatively similar. This shows that while (BS) CCSD and CCSD(T) are not usually considered to be multireference methods, they can be qualitatively correct for (spin-averaged) one-particle quantities, and thus for most conventional analyses of bonding.
The main problem with the energies obtained from the BS CC methods here is the lack of error cancellation for configurations with varying multireference character (namely, the error in the absolute BS-CCSD(T) energy for the HC and HFe structures can be quite different), rather than the complete failure of the CCSD and CCSD(T) methods. 

From the DMRG natural-orbital occupancy plot for the PS state (\autoref{fig:occ}(b)), we see that there are singly occupied orbitals associated with the Fe centers, but additionally zero, one, two and three orbitals with fractional occupancies between 0.2 and 0.8 (or between 1.2 and 1.8; marked by gray in \autoref{fig:occ}), respectively, for the HC, HS, HFe and \( \mathrm{HFe_2} \) structures. This clearly illustrates the trend of increasing multireference character, beyond spin-recoupling of the Fe centers, in this sequence of four structures.

\begin{figure*}[!htbp]
  \includegraphics[width=\linewidth]{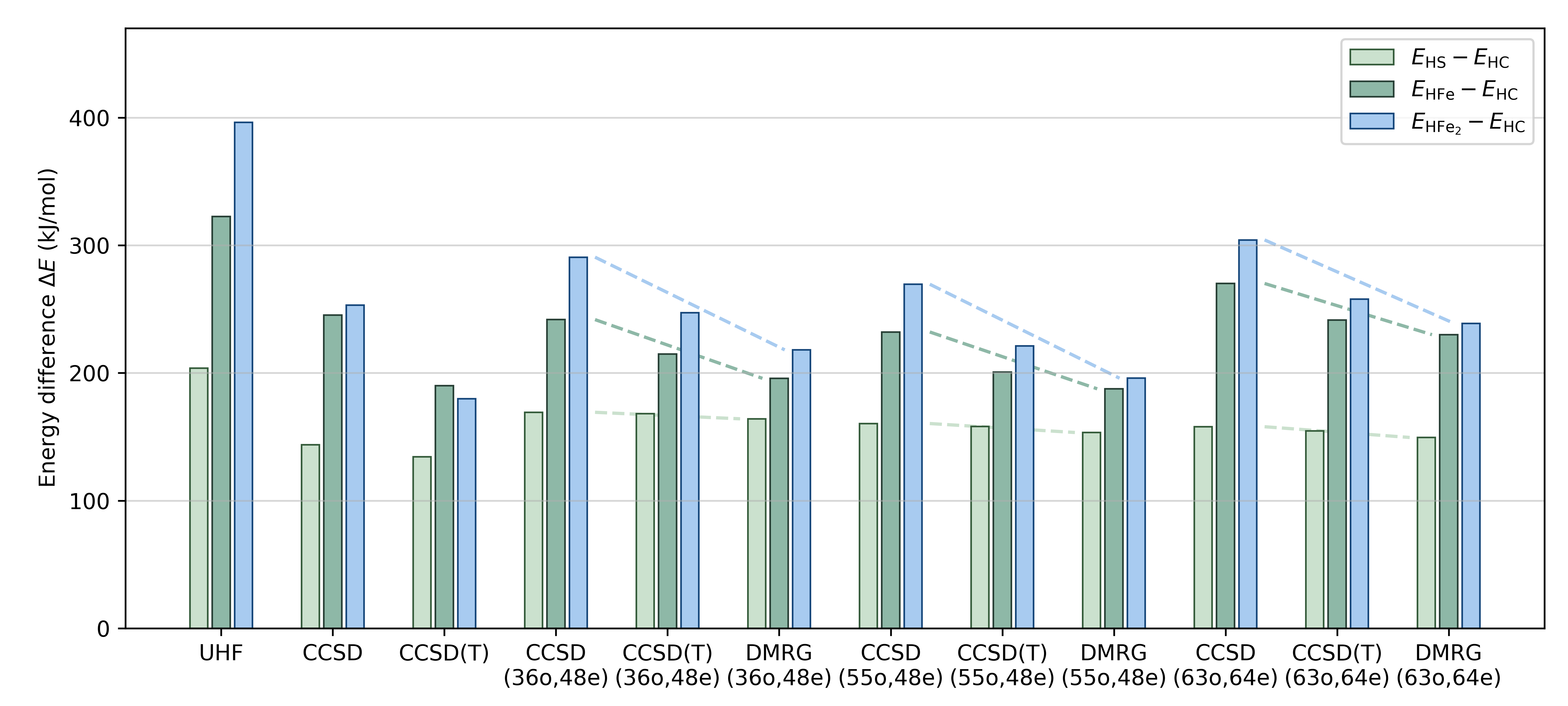}
  \caption{Relative single point energies for the protonated Fe dimers computed using UHF, CCSD, CCSD(T) and DMRG with the cc-pVDZ-DK basis, in the full orbital space and in the (36o, 48e) and (55o, 48e) active spaces. The energy of the HC structure is used as the reference. The trends in the (T) and DMRG correction for the relative energies are shown by dashed lines. }
  \label{fig:ener-dmrg}
\end{figure*}

\begin{table}[!htbp]
    \centering
    \caption{Comparison between UHF/CCSD(T) and DMRG energy corrections (in kJ/mol) for individual protonated structures computed using the cc-pVDZ-DK basis. The CCSD energy is used as the reference for CCSD(T) and DMRG energies. Note that the DMRG energies are computed for the PS singlet state while other energies are computed for the BS state. Energies and errors are listed for individual structures unless otherwise specified by ``(relative to HC)''.}
    \begin{tabular}{
        >{\centering\arraybackslash}p{3.4cm}
        >{\centering\arraybackslash}p{1.1cm}
        >{\centering\arraybackslash}p{1.1cm}
        >{\centering\arraybackslash}p{1.1cm}
        >{\centering\arraybackslash}p{1.1cm}
    }
    \hline\hline
    \multirow{2}{*}{Theory} & 
    \multicolumn{4}{c}{ Energy correction \( E - E_\mathrm{CCSD} \) } \\
    & HC & HS & HFe & \( \mathrm{HFe_2} \) \\
    \hline \\
    \multicolumn{5}{c}{active space (36o, 48e)} \\  \\
    CCSD (relative to HC) & 0.0 & 169.2 & 241.8 & 290.7 \\
    CCSD(T) & \(-3.4\) & \(-4.4\) & \(-30.4\) & \(-47.0\) \\
    DMRG (extrapolated) & \(-6.8\) & \(-11.9\) & \(-52.8\) & \(-79.4\) \\
    DMRG extrap. error & 0.0 & 0.0 & 0.1 & 0.2 \\
    \\
    \multicolumn{5}{c}{active space (55o, 48e)} \\  \\
    CCSD (relative to HC) & 0.0 & 160.4 & 232.1 & 269.5 \\
    CCSD(T) & \(-10.8\) & \(-13.2\) & \(-42.3\) & \(-59.2\) \\
    DMRG (extrapolated) & \(-16.5\)& \(-23.5\)& \(-61.0\) & \(-90.1\) \\
    DMRG extrap. error & 0.4 & 0.4 & 1.0 & 3.2 \\
    \\
    \multicolumn{5}{c}{active space (63o, 64e)} \\  \\
    CCSD (relative to HC) & 0.0 & 157.9 & 270.1 & 304.2 \\
    CCSD(T) & \(-18.3\) & \(-21.4\) & \(-46.9\) & \(-64.6\) \\
    DMRG (extrapolated) & \(-27.1\)& \(-35.5\)& \(-67.2\) & \(-92.6\) \\
    DMRG extrap. error & 0.5 & 0.5 & 1.2 & 2.8 \\
    \\
    \multicolumn{5}{c}{full orbital space} \\  \\
    CCSD (relative to HC) & 0.0 & 143.7 & 245.3 & 253.1 \\
    CCSD(T) & \(-214.0\) & \(-223.5\) & \(-269.2\) & \(-287.3\) \\
 \\
    \hline\hline
    \end{tabular}
    \label{tab:cc-dmrg}
\end{table}

In \autoref{fig:ener-dmrg} we compare the UHF, CCSD, CCSD(T) and DMRG energy differences for the four protonation states and in \autoref{tab:cc-dmrg} we compare the CCSD(T) and DMRG energy corrections to the CCSD relative energies for the individual structures. We see that within the  (36o, 48e), (55o, 48e) and (63o, 64e) active spaces, the (T) contribution to the energy difference between HC and \( \mathrm{HFe_2} \)
is 60\%, 66\% and 63\% of the contribution in the full orbital space, respectively. 
The largest estimated error for the extrapolated DMRG energy (\( 3\ \mathrm{kJ/mol} \)) illustrates that the DMRG energies are almost exact on the current scale of the relative energetics.
In all cases, the DMRG and (T) corrections to the CCSD relative energies are in the same direction, as indicated by the dashed lines in \autoref{fig:ener-dmrg}.

Based on the data above, we can estimate the errors in the CCSD(T) energies for the HC, HS, HFe and \( \mathrm{HFe_2} \) structures to be \( -3, -7, -22, \) and \( -32 \ \mathrm{kJ/mol} \) (from the 36o active space), \( -6, -10, -19, \) and \( -31 \ \mathrm{kJ/mol} \) (from the 55o active space), or \( -9, -14, -20, \) and \( -28 \ \mathrm{kJ/mol} \) (from the 63o active space), respectively.
The DMRG correction is thus quite small for the HC and HS structures, but larger for the HFe and \( \mathrm{HFe_2} \) structures, reflecting the multireference character of the Fe--H bond. 
From the DMRG bond dimension \( M = 5000 \) wave function in the 63o active space, we obtain a largest CSF weight for the four structures of 0.72, 0.66, 0.51 and 0.36, respectively. This confirms that the error in the CCSD(T) energy increases when the multireference character of the structure increases.

\begin{figure}[!htbp]
  \includegraphics[width=\linewidth]{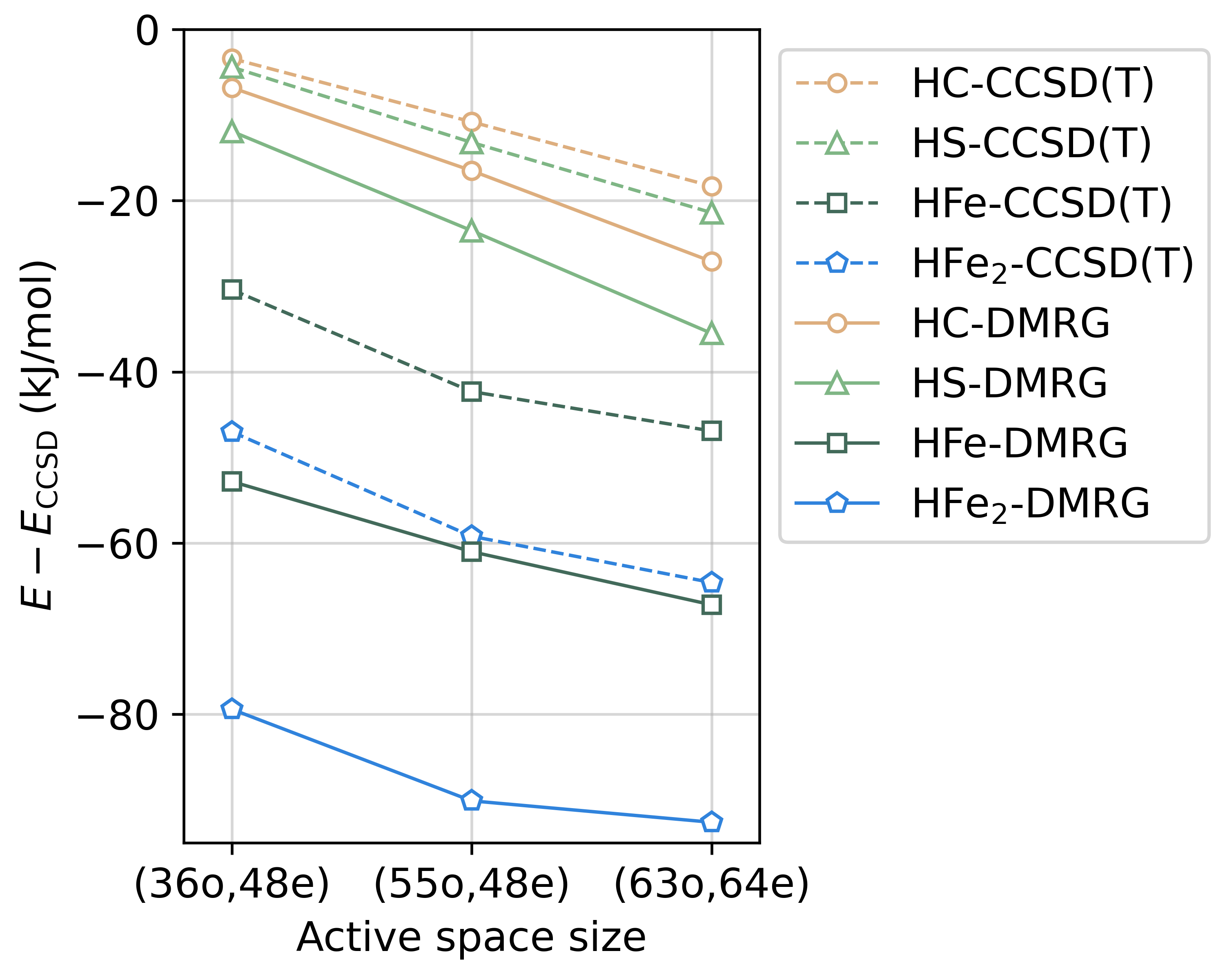}
  \caption{CCSD(T) and DMRG correlation energies, relative to CCSD, of the protonated Fe dimers computed for active spaces of different sizes (using UHF/CCSD natural orbitals) in the cc-pVDZ-DK basis set.}
  \label{fig:active}
\end{figure}

In Fig.~\ref{fig:active}, we show the trends in the correlation effects beyond CCSD as estimated by DMRG ($\Delta E_\mathrm{DMRG} = E_\mathrm{DMRG}-E_\mathrm{CCSD}$) and (T), for three active space sizes. The curves track each other,
justifying the possibility to use the composite energy formula. We use the DMRG results in the 63 orbital active space to correct for the missing multireference effect in the CCSD(T) energies. As discussed in the Methods section, we estimate the uncertainty of this correction for the relative energies as half of the correction for the HFe$_2$ structure (the largest correction), i.e. \( \pm 10\ \mathrm{kJ/mol} \).

\subsection{Basis-set correction and relativistic contribution}
\label{sec:basis-set}

\begin{figure}[!htbp]
  \includegraphics[width=\columnwidth]{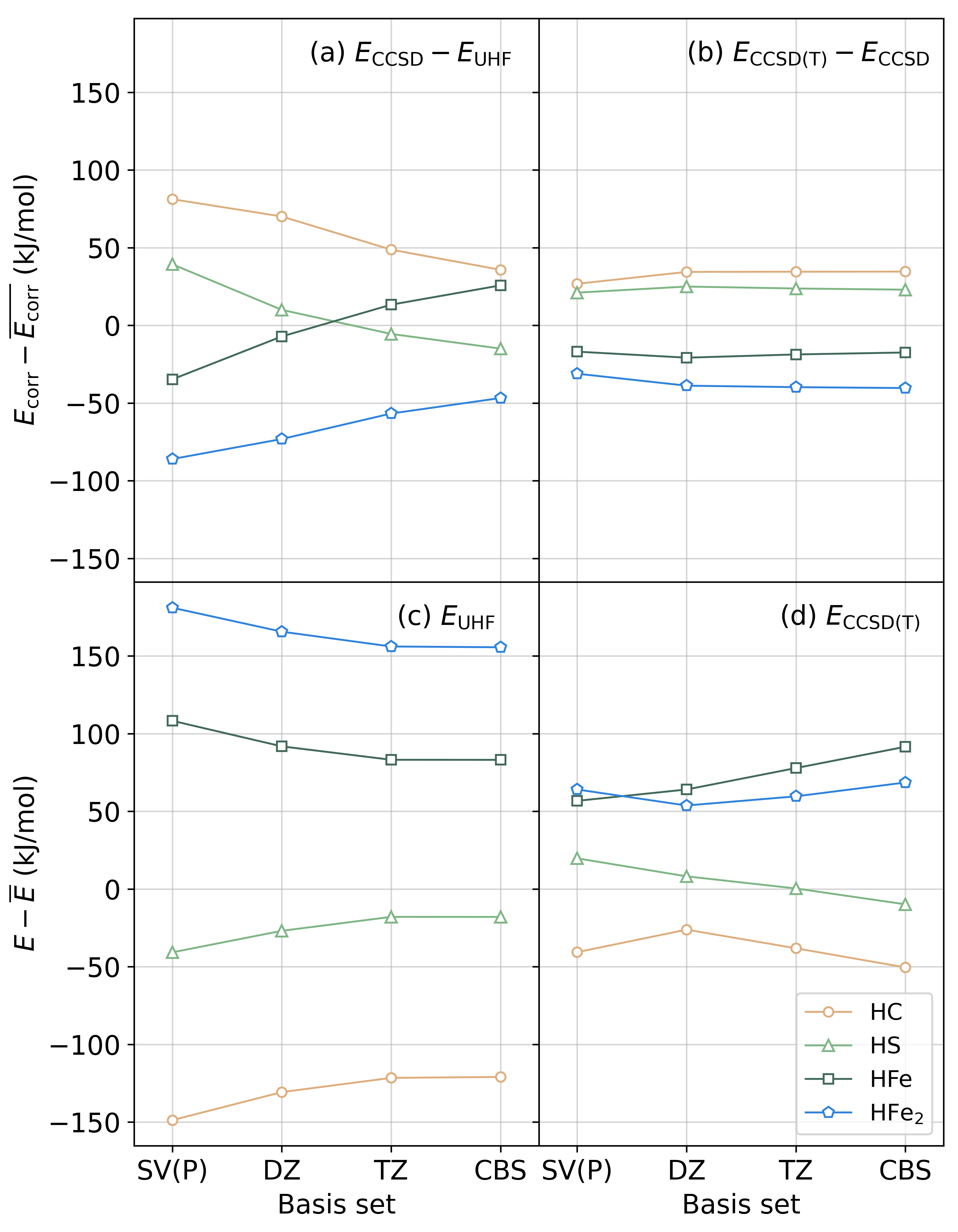}
  \caption{Trends in the energies of the protonated Fe dimers for (a) the CCSD correlation energies, (b) the (T) corrections, (c) the UHF energies and (d) the total CCSD(T) energies, as a function of basis set. For each basis set, the total or correlation energies are shifted by their average among the four structures. For UHF and CCSD(T) energies of the HC structure, an additional \( +100\ \mathrm{kJ/mol} \) shift is added for clarity.}
  \label{fig:basis}
\end{figure}

In order to study the basis-set effects on the CCSD(T) energies, we computed the UHF, MP2, CCSD and CCSD(T) energies for the four protonated structures using also larger basis sets. The results are listed in \autoref{tab:basis} and plotted in \autoref{fig:basis}.  We can see that the basis-set dependence of the UHF and correlation energies is very different, largely depending on whether the proton is bound to the metal or not. For the HC and HS structures, the UHF relative energies increase (become more positive) as the basis-set size increases, while the CCSD contributions decrease; for the HFe and \( \mathrm{HFe_2} \) structures, the trends are opposite. As a result, the basis-set dependence of the mean-field and correlation  energies partially cancel, and overall the total CCSD(T) relative energies change non-monotonically with increasing basis-set size. The UHF energies converge at the QZ level and the (T) corrections converge at the TZ level. Therefore, the CCSD(T) relative energy basis-set trend beyond TZ (bottom right panel, Fig.~\ref{fig:basis}) is dominated by the basis-set trend of the CCSD relative energies beyond TZ (top left panel, Fig.~\ref{fig:basis}).

Using the difference between the DZ/TZ- and TZ/QZ-CBS extrapolation energies computed at the MP2 level, we estimate the error at the DZ/TZ-CBS CCSD(T) level to be \( \pm 8\ \mathrm{kJ/mol} \) for the relative energy of the various structures. 

\begin{table}[!htbp]
    \centering
    \caption{The UHF, MP2 correlation, CCSD correlation and (T) correction energies computed using different basis sets. To better highlight trends in the relative energetics, we show the total or correlation energy averaged over the four structures for each basis set. This is then used as a reference energy. }
    \begin{tabular}{
        >{\centering\arraybackslash}p{1.9cm}
        >{\centering\arraybackslash}p{1.8cm}
        >{\centering\arraybackslash}p{1.0cm}
        >{\centering\arraybackslash}p{1.0cm}
        >{\centering\arraybackslash}p{1.0cm}
        >{\centering\arraybackslash}p{1.0cm}
    }
    \hline\hline
    \multirow{2}{*}{Basis} & \( E_{\mathrm{average}} \) & 
    \multicolumn{4}{c}{ \( E - E_\mathrm{average} \) (kJ/mol) } \\
    &(Hartree) & HC & HS & HFe & \( \mathrm{HFe_2} \) \\
    \hline \\
    \multicolumn{6}{c}{UHF} \\ \\
    def2-SV(P) & \(-4728.8928\)
                    & \(-248.7\) & \(-40.7\) & \(+108.3\) & \(+181.1\) \\
    cc-pVDZ-DK & \(-4733.8071\)
                    & \(-230.7\) & \(-26.9\) & \(+91.9\) & \(+165.7\) \\
    cc-pVTZ-DK & \(-4733.9598\)
                    & \(-221.5\) & \(-17.9\) & \(+83.2\) & \(+156.1\) \\
    cc-pVQZ-DK & \(-4734.0044\)
                    & \(-220.9\) & \(-17.9\) & \(+83.2\) & \(+155.6\) \\
    \\
    \multicolumn{6}{c}{UHF/MP2 (correlation energy)} \\ \\
    def2-SV(P) & \(-1.6831\)
                    & \(+49.5\) & \(+21.1\) & \(-12.7\) & \(-57.9\) \\
    cc-pVDZ-DK & \(-2.3403\)
                    & \(+27.9\) & \(-28.5\) & \(+36.4\) & \(-35.9\) \\
    cc-pVTZ-DK & \(-3.3457\)
                    & \(-0.6\) & \(-52.7\) & \(+66.2\) & \(-13.0\) \\
    cc-pVQZ-DK & \(-3.9214\)
                    & \(-5.0\) & \(-58.0\) & \(+71.6\) & \(-8.5\) \\
    TZ/QZ CBS  & \(-4.5001\)
                    & \(-9.5\) & \(-63.3\) & \(+77.0\) & \(-4.1\) \\
    DZ/TZ CBS  & \(-3.9565\)
                    & \(-17.9\) & \(-67.4\) & \(+84.3\) & \(+0.9\) \\
    \\
    \multicolumn{6}{c}{UHF/CCSD (correlation energy)} \\ \\
    def2-SV(P) & \(-1.8521\)
                    & \(+81.3\) & \(+39.4\) & \(-34.7\) & \(-86.0\) \\
    cc-pVDZ-DK & \(-2.4104\)
                    & \(+70.1\) & \(+10.0\) & \(-7.1\) & \(-73.1\) \\
    cc-pVTZ-DK & \(-3.1171\)
                    & \(+48.8\) & \(-5.5\) & \(+13.4\) & \(-56.7\) \\
    DZ/TZ CBS  & \(-3.5464\)
                    & \(+35.8\) & \(-14.9\) & \(+25.8\) & \(-46.7\) \\
    \\
    \multicolumn{6}{c}{UHF/CCSD(T) [(T) only]} \\ \\
    def2-SV(P) & \(-0.0701\) & \(+26.8\) & \(+21.1\) & \(-16.9\) & \(-31.1\) \\
    cc-pVDZ-DK & \(-0.0946\) & \(+34.5\) & \(+25.0\) & \(-20.7\) & \(-38.8\) \\
    cc-pVTZ-DK & \(-0.1491\) & \(+34.6\) & \(+23.8\) & \(-18.7\) & \(-39.7\) \\
    DZ/TZ CBS  & \(-0.1821\) & \(+34.7\) & \(+23.0\) & \(-17.4\) & \(-40.3\) \\
    \\
    \hline\hline
    \end{tabular}
    \label{tab:basis}
\end{table}

It is also interesting to break out the scalar relativistic contributions to the relative energies of the different structures, shown in \autoref{tab:other-corr}. For simplicity we used the def2-SV(P) basis set for a qualitative assessment. We see that the  scalar relativistic contribution is important for the relative energies of HFe and $\mathrm{HFe_2}$ [11–12 kJ/mol at the CCSD(T) level]. Relativistic effects are clearly necessary to describe differential bonding to Fe. 

\begin{table}[!htbp]
    \centering
    \caption{Scalar relativistic corrections (in kJ/mol) to the relative energies computed using different theories with the def2-SV(P) basis. The energy of the HC structure is used as the reference energy for all energies. \( \Delta E_{\mathrm{ref}} \) represents the energy difference with no relativistic corrections.}
    \begin{tabular}{
        >{\centering\arraybackslash}p{3.85cm}
        >{\centering\arraybackslash}p{1.0cm}
        >{\centering\arraybackslash}p{1.0cm}
        >{\centering\arraybackslash}p{1.0cm}
        >{\centering\arraybackslash}p{1.0cm}
    }
    \hline\hline
    \multirow{2}{*}{Theory} & 
    \multicolumn{4}{c}{ Energy difference \( E - E_\mathrm{HC} \) } \\
    & HC & HS & HFe & \( \mathrm{HFe_2} \) \\
    \hline \\
    \multicolumn{5}{c}{Relativistic: \( \Delta E_{\mathrm{X2C}} - \Delta E_{\mathrm{ref}}\) } \\ \\
    UHF            & 0.0 & \(-3.7\) & \(-21.5\) & \(-21.9\) \\
    UKS-TPSS       & 0.0 & \(-2.2\) & \(-8.7\)  & \(-7.7\) \\
    UKS-B3LYP      & 0.0 & \(-2.1\) & \(-10.1\) & \(-8.8\) \\
    \\
    UHF/CCSD    & 0.0 & \(-2.5\) & \(-13.7\) & \(-13.5\) \\
    UHF/CCSD(T) & 0.0 & \(-2.3\) & \(-11.9\) & \(-10.7\) \\
    \\
    \hline\hline
    \end{tabular}
    \label{tab:other-corr}
\end{table}

\subsection{Final composite energies and analysis}
\label{sec:total-ener}
 
\begin{figure*}[!htbp]
  \includegraphics[width=\linewidth]{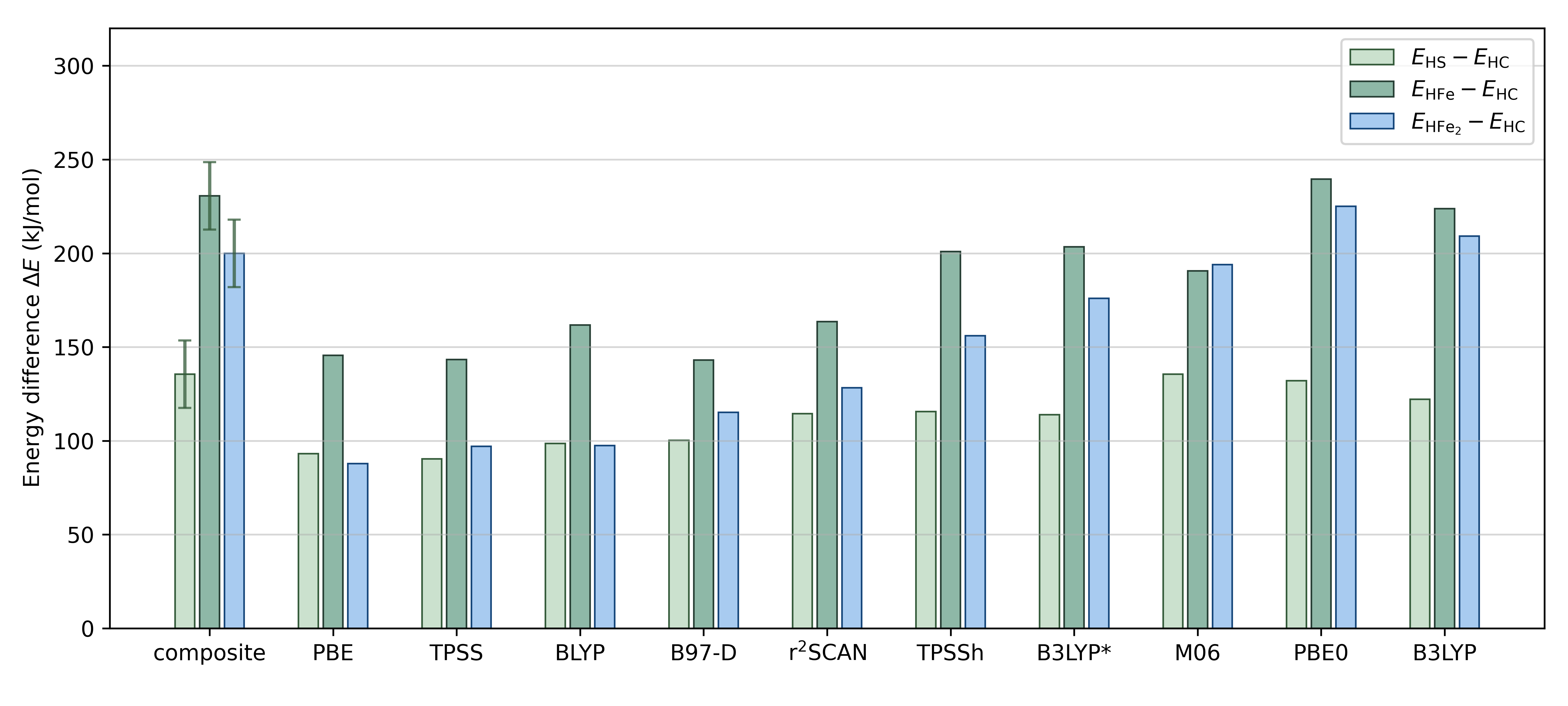}
  \caption{Comparison between the difference in single-point energy of the protonated Fe dimers computed using the composite CCSD(T)/DMRG approach and DFT with different functionals. The energy of the HC structure is used as the reference. The cc-pVQZ-DK basis set and CBS extrapolation results are used for mean-field and post-mean-field methods, respectively, unless otherwise specified. The uncertainty in the energy difference is shown as the error bar.}
  \label{fig:final}
\end{figure*}

In \autoref{tab:total} we summarize our final estimates for the relative energies of the four protonated structures obtained with the composite formula. We show the various contributions to the energy differences in \autoref{fig:final}. 
Overall, we find that \( E_{\mathrm{HC}} < E_{\mathrm{HS}} < E_{\mathrm{HFe_2}} < E_{\mathrm{HFe}} \). 

Both the basis-set and high-order correlation effects are important to obtain the correct qualitative ordering. While CCSD(T)/TZ may often be considered to produce reasonable results for the thermochemistry of small molecules, this is not the case for the Fe--S clusters: multireference effects beyond (T) and basis-set effects beyond TZ change the relative protonation energy of HC and \( \mathrm{HFe_2} \) by \(-19\) and \(+21\)~kJ/mol, respectively. As we needed to perform extrapolations to obtain both the multireference and basis-set corrections, our estimated uncertainty in these energies is \(\pm 10\) and \(\pm 8\)~kJ/mol, respectively. However, it must be stressed that our estimates of the uncertainties are quite crude. Interestingly, although the multireference and basis-set contributions are individually large, they have opposite signs. Consequently, the combined contribution is significantly smaller, and more closely resembles the raw CCSD(T)/TZ result.

\autoref{tab:total} and \autoref{fig:final} also include relative energies calculated with ten different DFT methods.
It can be seen that the BLYP, B97-D, r2SCAN, TPSSh, B3LYP*, B3LYP, and PBE0 functionals all obtain the correct qualitative ordering of the structures, while the TPSS, PBE, and M06 functionals do not. Out of the functionals with the correct ordering, the standard hybrid functionals B3LYP and PBE0 recover the composite method energetics to (approximately) within the estimated uncertainty in our composite results (18~kJ/mol, from adding the uncertainty in the multireference and basis-set correction) while the other functionals do not. Overall, there is a wide spread in the DFT predictions, for example, the range of the energy difference between HC and HS differs from our best estimate by 0--45 kJ/mol. 
The largest errors are found for the \( \mathrm{HFe_2} \) structure, where
the PBE functional gives an error in the relative protonation energy of 112~kJ/mol.
These effects are expected to be multiplied when there are multiple protons involved, as is the case for the $\mathrm{E_4}$ intermediate state of FeMoco. Our results are thus consistent with the large variation in protonation energies (hundreds of kJ/mol) observed when using different functionals to study multiply-protonated structures in the $\mathrm{E_4}$ intermediate~\cite{cao2019extremely}.   

\begin{table}[!htbp]
    \centering
    \caption{Relative single-point energies (in kJ/mol) for the four protonated Fe dimer structures computed by the composite method. The energy of the HC structure is used as the reference. UKS energies with different DFT functionals are also listed for comparison.
    The last two columns show the mean and max deviation
    in the energy difference \( E - E_\mathrm{HC} \) between the composite
    and UKS methods for each functional.}
    \begin{tabular}{
        >{\centering\arraybackslash}p{3.0cm}
        >{\centering\arraybackslash}p{0.8cm}
        >{\centering\arraybackslash}p{0.8cm}
        >{\centering\arraybackslash}p{0.8cm}
        >{\centering\arraybackslash}p{0.8cm}
        >{\centering\arraybackslash}p{0.8cm}
        >{\centering\arraybackslash}p{0.8cm}
    }
    \hline\hline
    \multirow{2}{*}{Correction/functional} & 
    \multicolumn{4}{c}{ Energy difference \( E - E_\mathrm{HC} \) } &
    \multicolumn{2}{c}{UKS deviation} \\
    & HC & HS & HFe & \( \mathrm{HFe_2} \) & mean & max \\
    \hline \\
    \multicolumn{7}{c}{ UHF/CCSD(T) (with X2C) } \\ \\
    uncorrected (cc-pVTZ-DK) & 0.0 & 138.6 & 216.1 & 197.9 && \\
    \\
    multireference correction & 0.0 & \(-5.2\) & \(-11.4\) & \(-19.2\) && \\
    basis-set correction & 0.0 & \(+2.1\) & \(+25.9\) & \(+21.2\) && \\
    \\
    total & 0.0 & 135.5	& 230.6 & 199.9 & 0.0 & 0.0 \\
    \\
    \multicolumn{7}{c}{ UKS (with X2C, DFT-D3 and cc-pVQZ-DK basis) } \\ \\
    PBE      & 0.0 &  93.1 & 145.6 &  87.8 & 79.8 & 112.1 \\
    BLYP     & 0.0 &  90.4 & 143.3 &  97.0 & 78.4 & 102.9 \\
    TPSS     & 0.0 &  98.6 & 161.8 &  97.4 & 69.4 & 102.5 \\
    B97-D    & 0.0 & 100.2 & 143.0 & 115.2 & 69.2 & 87.6 \\
r${}^2$SCAN   & 0.0 & 114.5 & 163.5 & 128.3& 53.2 & 71.6 \\
    TPSSh    & 0.0 & 115.6 & 200.9 & 156.0 & 31.2 & 43.9 \\
    B3LYP*   & 0.0 & 113.9 & 203.4 & 176.0 & 24.2 & 27.2 \\
    M06      & 0.0 & 135.5 & 190.6 & 194.0 & 15.3 & 40.0 \\
    PBE0     & 0.0 & 132.0 & 239.5 & 225.0 & 12.5 & 25.1 \\
    B3LYP    & 0.0 & 122.1 & 223.8 & 209.1 &  9.8 & 13.4 \\
   \\
    \hline\hline
    \end{tabular}
    \label{tab:total}
\end{table}

\section{Conclusions}

In this work, we studied the protonation energetics of a dimeric iron--sulfur cluster, as a simple model for the protonation of nitrogenase iron--sulfur clusters, such as the intermediates in FeMoco. Using a composite method based on CC and DMRG energies, we estimated the relative protonation energies of four representative structures (protonated on C, S, Fe or bridging two Fe) in the multireference and basis-set limits.
We found that both multireference and basis-set effects are extremely important to capturing the correct energy ordering. Importantly, even though we are studying the seemingly simple process of adding a single proton to the cluster, basis-set effects beyond triple-zeta, and correlation effects beyond (perturbative) triples, contribute about 20 kJ/mol to the relative energies (although the contributions have opposite signs). This highlights the challenge of computing accurate energetics for even larger clusters. 

The current work relies on a number of extrapolations to obtain the basis-set and correlation-effect limits. These extrapolations, as well as the crude estimates of the errors associated with them, are not entirely satisfactory. While some of these steps could be removed by performing more demanding computations, it may be challenging to scale such a strategy to the larger iron--sulfur clusters. In particular, while perturbative triples formed a reasonable starting point for the relative energetics in this cluster, it is unclear whether this will be the case in larger iron--sulfur clusters. The density functionals that we examined yielded a wide range of predictions, from qualitatively incorrect results, to results compatible with our estimates, with the hybrid B3LYP functional giving the best results. 
The different behaviors of the functionals highlights the well-known importance of tailoring the functional in challenging transition metal problems. 
Benchmark energetics, such as those from this work, thus serve as a starting point for choosing appropriate functionals to explore the chemistry of larger Fe--S clusters.

\section*{Supplementary Material}
Cluster geometries, DMRG energy extrapolations performed in the three active spaces, and the figures of the CCSD natural orbitals for constructing the active space.

\begin{acknowledgments}
Work at Caltech was supported by the 
Center for Molecular Magnetic Quantum Materials, an Energy Frontier Research Center funded by the U.S. Department of Energy, Office of Science, Basic Energy Sciences under award no. DE-SC0019330. 
The computations were conducted at the Resnick High Performance Computing Center, a facility supported by the Resnick Sustainability Institute at the California Institute of Technology.
Work at Lund University was supported by grants from the Swedish research council (projects 2018-05003 and 2022-04978). The computations were performed on computer resources provided by the Swedish National Infrastructure for Computing (SNIC) at Lunarc at Lund University and HPC2N at Umeå University, partially funded by the Swedish Research Council (grant 2018-05973).
\end{acknowledgments}

\section*{Conflict of Interest}
The authors have no conflicts to disclose.

\section*{Data Availability}
The data presented in this work can be reproduced using the open-source code \textsc{PySCF} 2.0.1\cite{sun2018pyscf,sun2020recent} and \textsc{block2} 0.5.1.\cite{zhai2021low} The reference input and output files can be found in the GitHub repository at \url{https://github.com/hczhai/fe-dimer-data}.

\bibliography{main}

\providecommand{\latin}[1]{#1}
\makeatletter
\providecommand{\doi}
  {\begingroup\let\do\@makeother\dospecials
  \catcode`\{=1 \catcode`\}=2 \doi@aux}
\providecommand{\doi@aux}[1]{\endgroup\texttt{#1}}
\makeatother
\providecommand*\mcitethebibliography{\thebibliography}
\csname @ifundefined\endcsname{endmcitethebibliography}
  {\let\endmcitethebibliography\endthebibliography}{}
\begin{mcitethebibliography}{68}
\providecommand*\natexlab[1]{#1}
\providecommand*\mciteSetBstSublistMode[1]{}
\providecommand*\mciteSetBstMaxWidthForm[2]{}
\providecommand*\mciteBstWouldAddEndPuncttrue
  {\def\EndOfBibitem{\unskip.}}
\providecommand*\mciteBstWouldAddEndPunctfalse
  {\let\EndOfBibitem\relax}
\providecommand*\mciteSetBstMidEndSepPunct[3]{}
\providecommand*\mciteSetBstSublistLabelBeginEnd[3]{}
\providecommand*\EndOfBibitem{}
\mciteSetBstSublistMode{f}
\mciteSetBstMaxWidthForm{subitem}{(\alph{mcitesubitemcount})}
\mciteSetBstSublistLabelBeginEnd
  {\mcitemaxwidthsubitemform\space}
  {\relax}
  {\relax}

\bibitem[Beinert \latin{et~al.}(1997)Beinert, Holm, and Munck]{beinert1997iron}
Beinert,~H.; Holm,~R.~H.; Munck,~E. Iron-sulfur clusters: nature's modular,
  multipurpose structures. \emph{Science} \textbf{1997}, \emph{277},
  653--659\relax
\mciteBstWouldAddEndPuncttrue
\mciteSetBstMidEndSepPunct{\mcitedefaultmidpunct}
{\mcitedefaultendpunct}{\mcitedefaultseppunct}\relax
\EndOfBibitem
\bibitem[Tanifuji and Ohki(2020)Tanifuji, and Ohki]{tanifuji2020metal}
Tanifuji,~K.; Ohki,~Y. Metal--sulfur compounds in N2 reduction and
  nitrogenase-related chemistry. \emph{Chemical reviews} \textbf{2020},
  \emph{120}, 5194--5251\relax
\mciteBstWouldAddEndPuncttrue
\mciteSetBstMidEndSepPunct{\mcitedefaultmidpunct}
{\mcitedefaultendpunct}{\mcitedefaultseppunct}\relax
\EndOfBibitem
\bibitem[Seefeldt \latin{et~al.}(2020)Seefeldt, Yang, Lukoyanov, Harris, Dean,
  Raugei, and Hoffman]{seefeldt2020reduction}
Seefeldt,~L.~C.; Yang,~Z.-Y.; Lukoyanov,~D.~A.; Harris,~D.~F.; Dean,~D.~R.;
  Raugei,~S.; Hoffman,~B.~M. Reduction of substrates by nitrogenases.
  \emph{Chemical reviews} \textbf{2020}, \emph{120}, 5082--5106\relax
\mciteBstWouldAddEndPuncttrue
\mciteSetBstMidEndSepPunct{\mcitedefaultmidpunct}
{\mcitedefaultendpunct}{\mcitedefaultseppunct}\relax
\EndOfBibitem
\bibitem[Rutledge and Tezcan(2020)Rutledge, and Tezcan]{rutledge2020electron}
Rutledge,~H.~L.; Tezcan,~F.~A. Electron transfer in nitrogenase. \emph{Chemical
  reviews} \textbf{2020}, \emph{120}, 5158--5193\relax
\mciteBstWouldAddEndPuncttrue
\mciteSetBstMidEndSepPunct{\mcitedefaultmidpunct}
{\mcitedefaultendpunct}{\mcitedefaultseppunct}\relax
\EndOfBibitem
\bibitem[Kirn and Rees(1992)Kirn, and Rees]{kirn1992crystallographic}
Kirn,~J.; Rees,~D. Crystallographic structure and functional implications of
  the nitrogenase molybdenum--iron protein from Azotobacter vinelandii.
  \emph{Nature} \textbf{1992}, \emph{360}, 553--560\relax
\mciteBstWouldAddEndPuncttrue
\mciteSetBstMidEndSepPunct{\mcitedefaultmidpunct}
{\mcitedefaultendpunct}{\mcitedefaultseppunct}\relax
\EndOfBibitem
\bibitem[Peters \latin{et~al.}(1997)Peters, Stowell, Soltis, Finnegan, Johnson,
  and Rees]{peters1997redox}
Peters,~J.~W.; Stowell,~M.~H.; Soltis,~S.~M.; Finnegan,~M.~G.; Johnson,~M.~K.;
  Rees,~D.~C. Redox-dependent structural changes in the nitrogenase P-cluster.
  \emph{Biochemistry} \textbf{1997}, \emph{36}, 1181--1187\relax
\mciteBstWouldAddEndPuncttrue
\mciteSetBstMidEndSepPunct{\mcitedefaultmidpunct}
{\mcitedefaultendpunct}{\mcitedefaultseppunct}\relax
\EndOfBibitem
\bibitem[Einsle \latin{et~al.}(2002)Einsle, Tezcan, Andrade, Schmid, Yoshida,
  Howard, and Rees]{einsle2002nitrogenase}
Einsle,~O.; Tezcan,~F.~A.; Andrade,~S.~L.; Schmid,~B.; Yoshida,~M.;
  Howard,~J.~B.; Rees,~D.~C. Nitrogenase MoFe-protein at 1.16 {\AA} resolution:
  a central ligand in the FeMo-cofactor. \emph{Science} \textbf{2002},
  \emph{297}, 1696--1700\relax
\mciteBstWouldAddEndPuncttrue
\mciteSetBstMidEndSepPunct{\mcitedefaultmidpunct}
{\mcitedefaultendpunct}{\mcitedefaultseppunct}\relax
\EndOfBibitem
\bibitem[Spatzal \latin{et~al.}(2011)Spatzal, Aksoyoglu, Zhang, Andrade,
  Schleicher, Weber, Rees, and Einsle]{spatzal2011evidence}
Spatzal,~T.; Aksoyoglu,~M.; Zhang,~L.; Andrade,~S.~L.; Schleicher,~E.;
  Weber,~S.; Rees,~D.~C.; Einsle,~O. Evidence for interstitial carbon in
  nitrogenase FeMo cofactor. \emph{Science} \textbf{2011}, \emph{334},
  940--940\relax
\mciteBstWouldAddEndPuncttrue
\mciteSetBstMidEndSepPunct{\mcitedefaultmidpunct}
{\mcitedefaultendpunct}{\mcitedefaultseppunct}\relax
\EndOfBibitem
\bibitem[Lancaster \latin{et~al.}(2011)Lancaster, Roemelt, Ettenhuber, Hu,
  Ribbe, Neese, Bergmann, and DeBeer]{lancaster2011x}
Lancaster,~K.~M.; Roemelt,~M.; Ettenhuber,~P.; Hu,~Y.; Ribbe,~M.~W.; Neese,~F.;
  Bergmann,~U.; DeBeer,~S. X-ray emission spectroscopy evidences a central
  carbon in the nitrogenase iron-molybdenum cofactor. \emph{Science}
  \textbf{2011}, \emph{334}, 974--977\relax
\mciteBstWouldAddEndPuncttrue
\mciteSetBstMidEndSepPunct{\mcitedefaultmidpunct}
{\mcitedefaultendpunct}{\mcitedefaultseppunct}\relax
\EndOfBibitem
\bibitem[Jafari \latin{et~al.}(2022)Jafari, Tavares~Santos, Bergmann, Irani,
  and Ryde]{jafari2022benchmark}
Jafari,~S.; Tavares~Santos,~Y.~A.; Bergmann,~J.; Irani,~M.; Ryde,~U. Benchmark
  Study of Redox Potential Calculations for Iron--Sulfur Clusters in Proteins.
  \emph{Inorganic chemistry} \textbf{2022}, \emph{61}, 5991--6007\relax
\mciteBstWouldAddEndPuncttrue
\mciteSetBstMidEndSepPunct{\mcitedefaultmidpunct}
{\mcitedefaultendpunct}{\mcitedefaultseppunct}\relax
\EndOfBibitem
\bibitem[Benediktsson and Bjornsson(2022)Benediktsson, and
  Bjornsson]{benediktsson2022analysis}
Benediktsson,~B.; Bjornsson,~R. Analysis of the Geometric and Electronic
  Structure of Spin-Coupled Iron--Sulfur Dimers with Broken-Symmetry DFT:
  Implications for FeMoco. \emph{Journal of chemical theory and computation}
  \textbf{2022}, \emph{18}, 1437--1457\relax
\mciteBstWouldAddEndPuncttrue
\mciteSetBstMidEndSepPunct{\mcitedefaultmidpunct}
{\mcitedefaultendpunct}{\mcitedefaultseppunct}\relax
\EndOfBibitem
\bibitem[Hoeke \latin{et~al.}(2019)Hoeke, Tociu, Case, Seefeldt, Raugei, and
  Hoffman]{hoeke2019high}
Hoeke,~V.; Tociu,~L.; Case,~D.~A.; Seefeldt,~L.~C.; Raugei,~S.; Hoffman,~B.~M.
  High-resolution ENDOR spectroscopy combined with quantum chemical
  calculations reveals the structure of nitrogenase Janus intermediate E4 (4H).
  \emph{Journal of the American Chemical Society} \textbf{2019}, \emph{141},
  11984--11996\relax
\mciteBstWouldAddEndPuncttrue
\mciteSetBstMidEndSepPunct{\mcitedefaultmidpunct}
{\mcitedefaultendpunct}{\mcitedefaultseppunct}\relax
\EndOfBibitem
\bibitem[Raugei \latin{et~al.}(2018)Raugei, Seefeldt, and Hoffman]{Hoffman:18}
Raugei,~S.; Seefeldt,~L.~C.; Hoffman,~B.~M. Critical computational analysis
  illuminates the reductive-elimination mechanism that activates nitrogenase
  for N 2 reduction. \emph{Proceedings of the National Academy of Sciences}
  \textbf{2018}, \emph{115}, E10521--E10530\relax
\mciteBstWouldAddEndPuncttrue
\mciteSetBstMidEndSepPunct{\mcitedefaultmidpunct}
{\mcitedefaultendpunct}{\mcitedefaultseppunct}\relax
\EndOfBibitem
\bibitem[Rohde \latin{et~al.}(2018)Rohde, Sippel, Trncik, Andrade, and
  Einsle]{Einsle:14}
Rohde,~M.; Sippel,~D.; Trncik,~C.; Andrade,~S. L.~A.; Einsle,~O. The Critical
  E$_4$ State of Nitrogenase Catalysis. \emph{Biochemistry} \textbf{2018},
  \emph{57}, 5497--5504, doi: 10.1021/acs.biochem.8b00509\relax
\mciteBstWouldAddEndPuncttrue
\mciteSetBstMidEndSepPunct{\mcitedefaultmidpunct}
{\mcitedefaultendpunct}{\mcitedefaultseppunct}\relax
\EndOfBibitem
\bibitem[Siegbahn(2016)]{Siegbahn:16}
Siegbahn,~P. E.~M. Model Calculations Suggest that the Central Carbon in the
  FeMo-Cofactor of Nitrogenase Becomes Protonated in the Process of Nitrogen
  Fixation. \emph{Journal of the American Chemical Society} \textbf{2016},
  \emph{138}, 10485--10495\relax
\mciteBstWouldAddEndPuncttrue
\mciteSetBstMidEndSepPunct{\mcitedefaultmidpunct}
{\mcitedefaultendpunct}{\mcitedefaultseppunct}\relax
\EndOfBibitem
\bibitem[Dance(2020)]{Dance:20}
Dance,~I. Computational Investigations of the Chemical Mechanism of the Enzyme
  Nitrogenase. \emph{ChemBioChem} \textbf{2020}, \emph{21}, 1671--1709\relax
\mciteBstWouldAddEndPuncttrue
\mciteSetBstMidEndSepPunct{\mcitedefaultmidpunct}
{\mcitedefaultendpunct}{\mcitedefaultseppunct}\relax
\EndOfBibitem
\bibitem[Cao \latin{et~al.}(2018)Cao, Caldararu, and Ryde]{cao2018protonation}
Cao,~L.; Caldararu,~O.; Ryde,~U. Protonation and reduction of the FeMo cluster
  in nitrogenase studied by quantum mechanics/molecular mechanics (QM/MM)
  calculations. \emph{Journal of chemical theory and computation}
  \textbf{2018}, \emph{14}, 6653--6678\relax
\mciteBstWouldAddEndPuncttrue
\mciteSetBstMidEndSepPunct{\mcitedefaultmidpunct}
{\mcitedefaultendpunct}{\mcitedefaultseppunct}\relax
\EndOfBibitem
\bibitem[Cao and Ryde(2020)Cao, and Ryde]{cao2020structure}
Cao,~L.; Ryde,~U. What Is the Structure of the E4 Intermediate in Nitrogenase?
  \emph{Journal of chemical theory and computation} \textbf{2020}, \emph{16},
  1936--1952\relax
\mciteBstWouldAddEndPuncttrue
\mciteSetBstMidEndSepPunct{\mcitedefaultmidpunct}
{\mcitedefaultendpunct}{\mcitedefaultseppunct}\relax
\EndOfBibitem
\bibitem[Thorhallsson \latin{et~al.}(2019)Thorhallsson, Benediktsson, and
  Bjornsson]{thorhallsson2019model}
Thorhallsson,~A.~T.; Benediktsson,~B.; Bjornsson,~R. A model for dinitrogen
  binding in the E 4 state of nitrogenase. \emph{Chemical science}
  \textbf{2019}, \emph{10}, 11110--11124\relax
\mciteBstWouldAddEndPuncttrue
\mciteSetBstMidEndSepPunct{\mcitedefaultmidpunct}
{\mcitedefaultendpunct}{\mcitedefaultseppunct}\relax
\EndOfBibitem
\bibitem[Lovell \latin{et~al.}(2001)Lovell, Li, Liu, Case, and
  Noodleman]{Noodleman:21}
Lovell,~T.; Li,~J.; Liu,~T.; Case,~D.~A.; Noodleman,~L. FeMo Cofactor of
  Nitrogenase: A Density Functional Study of States $\mathrm{M^N}$,
  $\mathrm{M^{OX}}$, $\mathrm{M^R}$, and $\mathrm{M^I}$. \emph{Journal of
  American Chemical Society} \textbf{2001}, \emph{123}, 12392--12410\relax
\mciteBstWouldAddEndPuncttrue
\mciteSetBstMidEndSepPunct{\mcitedefaultmidpunct}
{\mcitedefaultendpunct}{\mcitedefaultseppunct}\relax
\EndOfBibitem
\bibitem[Cao and Ryde(2018)Cao, and Ryde]{Cao:18-bs}
Cao,~L.; Ryde,~U. Influence of the protein and DFT method on the
  broken-symmetry and spin states in nitrogenase. \emph{International Journal
  of Quantum Chemistry} \textbf{2018}, \emph{118}, e25627\relax
\mciteBstWouldAddEndPuncttrue
\mciteSetBstMidEndSepPunct{\mcitedefaultmidpunct}
{\mcitedefaultendpunct}{\mcitedefaultseppunct}\relax
\EndOfBibitem
\bibitem[Hohenberg and Kohn(1964)Hohenberg, and
  Kohn]{hohenberg1964inhomogeneous}
Hohenberg,~P.; Kohn,~W. Inhomogeneous electron gas. \emph{Physical Review}
  \textbf{1964}, \emph{136}, B864\relax
\mciteBstWouldAddEndPuncttrue
\mciteSetBstMidEndSepPunct{\mcitedefaultmidpunct}
{\mcitedefaultendpunct}{\mcitedefaultseppunct}\relax
\EndOfBibitem
\bibitem[Cao and Ryde(2019)Cao, and Ryde]{cao2019extremely}
Cao,~L.; Ryde,~U. Extremely large differences in DFT energies for nitrogenase
  models. \emph{Physical Chemistry Chemical Physics} \textbf{2019}, \emph{21},
  2480--2488\relax
\mciteBstWouldAddEndPuncttrue
\mciteSetBstMidEndSepPunct{\mcitedefaultmidpunct}
{\mcitedefaultendpunct}{\mcitedefaultseppunct}\relax
\EndOfBibitem
\bibitem[Wei and Siegbahn(2021)Wei, and Siegbahn]{wei2021active}
Wei,~W.-J.; Siegbahn,~P.~E. The active E4 structure of nitrogenase studied with
  different DFT functionals. \emph{Journal of computational chemistry}
  \textbf{2021}, \emph{42}, 81--85\relax
\mciteBstWouldAddEndPuncttrue
\mciteSetBstMidEndSepPunct{\mcitedefaultmidpunct}
{\mcitedefaultendpunct}{\mcitedefaultseppunct}\relax
\EndOfBibitem
\bibitem[White(1992)]{white1992density}
White,~S.~R. Density matrix formulation for quantum renormalization groups.
  \emph{Physical review letters} \textbf{1992}, \emph{69}, 2863\relax
\mciteBstWouldAddEndPuncttrue
\mciteSetBstMidEndSepPunct{\mcitedefaultmidpunct}
{\mcitedefaultendpunct}{\mcitedefaultseppunct}\relax
\EndOfBibitem
\bibitem[White(1993)]{white1993density}
White,~S.~R. Density-matrix algorithms for quantum renormalization groups.
  \emph{Physical Review B} \textbf{1993}, \emph{48}, 10345\relax
\mciteBstWouldAddEndPuncttrue
\mciteSetBstMidEndSepPunct{\mcitedefaultmidpunct}
{\mcitedefaultendpunct}{\mcitedefaultseppunct}\relax
\EndOfBibitem
\bibitem[Chan and Head-Gordon(2002)Chan, and Head-Gordon]{chan2002highly}
Chan,~G. K.-L.; Head-Gordon,~M. Highly correlated calculations with a
  polynomial cost algorithm: A study of the density matrix renormalization
  group. \emph{The Journal of chemical physics} \textbf{2002}, \emph{116},
  4462--4476\relax
\mciteBstWouldAddEndPuncttrue
\mciteSetBstMidEndSepPunct{\mcitedefaultmidpunct}
{\mcitedefaultendpunct}{\mcitedefaultseppunct}\relax
\EndOfBibitem
\bibitem[Chan and Sharma(2011)Chan, and Sharma]{chan2011density}
Chan,~G. K.-L.; Sharma,~S. The density matrix renormalization group in quantum
  chemistry. \emph{Annual review of physical chemistry} \textbf{2011},
  \emph{62}, 465--481\relax
\mciteBstWouldAddEndPuncttrue
\mciteSetBstMidEndSepPunct{\mcitedefaultmidpunct}
{\mcitedefaultendpunct}{\mcitedefaultseppunct}\relax
\EndOfBibitem
\bibitem[Mitrushenkov \latin{et~al.}(2001)Mitrushenkov, Fano, Ortolani,
  Linguerri, and Palmieri]{mitrushenkov2001quantum}
Mitrushenkov,~A.~O.; Fano,~G.; Ortolani,~F.; Linguerri,~R.; Palmieri,~P.
  Quantum chemistry using the density matrix renormalization group. \emph{The
  Journal of Chemical Physics} \textbf{2001}, \emph{115}, 6815--6821\relax
\mciteBstWouldAddEndPuncttrue
\mciteSetBstMidEndSepPunct{\mcitedefaultmidpunct}
{\mcitedefaultendpunct}{\mcitedefaultseppunct}\relax
\EndOfBibitem
\bibitem[Chan \latin{et~al.}(2016)Chan, Keselman, Nakatani, Li, and
  White]{chan2016matrix}
Chan,~G. K.-L.; Keselman,~A.; Nakatani,~N.; Li,~Z.; White,~S.~R. Matrix product
  operators, matrix product states, and ab initio density matrix
  renormalization group algorithms. \emph{The Journal of chemical physics}
  \textbf{2016}, \emph{145}, 014102\relax
\mciteBstWouldAddEndPuncttrue
\mciteSetBstMidEndSepPunct{\mcitedefaultmidpunct}
{\mcitedefaultendpunct}{\mcitedefaultseppunct}\relax
\EndOfBibitem
\bibitem[Sharma and Chan(2012)Sharma, and Chan]{sharma2012spin}
Sharma,~S.; Chan,~G. K.-L. Spin-adapted density matrix renormalization group
  algorithms for quantum chemistry. \emph{The Journal of chemical physics}
  \textbf{2012}, \emph{136}, 124121\relax
\mciteBstWouldAddEndPuncttrue
\mciteSetBstMidEndSepPunct{\mcitedefaultmidpunct}
{\mcitedefaultendpunct}{\mcitedefaultseppunct}\relax
\EndOfBibitem
\bibitem[Olivares-Amaya \latin{et~al.}(2015)Olivares-Amaya, Hu, Nakatani,
  Sharma, Yang, and Chan]{olivares2015ab}
Olivares-Amaya,~R.; Hu,~W.; Nakatani,~N.; Sharma,~S.; Yang,~J.; Chan,~G. K.-L.
  The ab-initio density matrix renormalization group in practice. \emph{The
  Journal of chemical physics} \textbf{2015}, \emph{142}, 034102\relax
\mciteBstWouldAddEndPuncttrue
\mciteSetBstMidEndSepPunct{\mcitedefaultmidpunct}
{\mcitedefaultendpunct}{\mcitedefaultseppunct}\relax
\EndOfBibitem
\bibitem[Wouters and Van~Neck(2014)Wouters, and Van~Neck]{wouters2014density}
Wouters,~S.; Van~Neck,~D. The density matrix renormalization group for ab
  initio quantum chemistry. \emph{The European Physical Journal D}
  \textbf{2014}, \emph{68}, 272\relax
\mciteBstWouldAddEndPuncttrue
\mciteSetBstMidEndSepPunct{\mcitedefaultmidpunct}
{\mcitedefaultendpunct}{\mcitedefaultseppunct}\relax
\EndOfBibitem
\bibitem[Keller \latin{et~al.}(2015)Keller, Dolfi, Troyer, and
  Reiher]{keller2015efficient}
Keller,~S.; Dolfi,~M.; Troyer,~M.; Reiher,~M. An efficient matrix product
  operator representation of the quantum chemical Hamiltonian. \emph{The
  Journal of chemical physics} \textbf{2015}, \emph{143}, 244118\relax
\mciteBstWouldAddEndPuncttrue
\mciteSetBstMidEndSepPunct{\mcitedefaultmidpunct}
{\mcitedefaultendpunct}{\mcitedefaultseppunct}\relax
\EndOfBibitem
\bibitem[Baiardi and Reiher(2020)Baiardi, and Reiher]{baiardi2020density}
Baiardi,~A.; Reiher,~M. The density matrix renormalization group in chemistry
  and molecular physics: Recent developments and new challenges. \emph{The
  Journal of Chemical Physics} \textbf{2020}, \emph{152}, 040903\relax
\mciteBstWouldAddEndPuncttrue
\mciteSetBstMidEndSepPunct{\mcitedefaultmidpunct}
{\mcitedefaultendpunct}{\mcitedefaultseppunct}\relax
\EndOfBibitem
\bibitem[Brabec \latin{et~al.}(2021)Brabec, Brandejs, Kowalski, Xantheas,
  Legeza, and Veis]{brabec2021massively}
Brabec,~J.; Brandejs,~J.; Kowalski,~K.; Xantheas,~S.; Legeza,~{\"O}.; Veis,~L.
  Massively parallel quantum chemical density matrix renormalization group
  method. \emph{Journal of Computational Chemistry} \textbf{2021}, \emph{42},
  534--544\relax
\mciteBstWouldAddEndPuncttrue
\mciteSetBstMidEndSepPunct{\mcitedefaultmidpunct}
{\mcitedefaultendpunct}{\mcitedefaultseppunct}\relax
\EndOfBibitem
\bibitem[Zhai and Chan(2021)Zhai, and Chan]{zhai2021low}
Zhai,~H.; Chan,~G. K.-L. Low communication high performance ab initio density
  matrix renormalization group algorithms. \emph{The Journal of Chemical
  Physics} \textbf{2021}, \emph{154}, 224116\relax
\mciteBstWouldAddEndPuncttrue
\mciteSetBstMidEndSepPunct{\mcitedefaultmidpunct}
{\mcitedefaultendpunct}{\mcitedefaultseppunct}\relax
\EndOfBibitem
\bibitem[Sharma \latin{et~al.}(2014)Sharma, Sivalingam, Neese, and
  Chan]{sharma2014low}
Sharma,~S.; Sivalingam,~K.; Neese,~F.; Chan,~G.~K. Low-energy spectrum of
  iron--sulfur clusters directly from many-particle quantum mechanics.
  \emph{Nature chemistry} \textbf{2014}, \emph{6}, 927--933\relax
\mciteBstWouldAddEndPuncttrue
\mciteSetBstMidEndSepPunct{\mcitedefaultmidpunct}
{\mcitedefaultendpunct}{\mcitedefaultseppunct}\relax
\EndOfBibitem
\bibitem[Li \latin{et~al.}(2019)Li, Guo, Sun, and Chan]{li2019electronic}
Li,~Z.; Guo,~S.; Sun,~Q.; Chan,~G.~K. Electronic landscape of the P-cluster of
  nitrogenase as revealed through many-electron quantum wavefunction
  simulations. \emph{Nature chemistry} \textbf{2019}, \emph{11},
  1026--1033\relax
\mciteBstWouldAddEndPuncttrue
\mciteSetBstMidEndSepPunct{\mcitedefaultmidpunct}
{\mcitedefaultendpunct}{\mcitedefaultseppunct}\relax
\EndOfBibitem
\bibitem[Li \latin{et~al.}(2019)Li, Li, Dattani, Umrigar, and Chan]{li2019the}
Li,~Z.; Li,~J.; Dattani,~N.~S.; Umrigar,~C.; Chan,~G. K.-L. The electronic
  complexity of the ground-state of the FeMo cofactor of nitrogenase as
  relevant to quantum simulations. \emph{The Journal of chemical physics}
  \textbf{2019}, \emph{150}, 024302\relax
\mciteBstWouldAddEndPuncttrue
\mciteSetBstMidEndSepPunct{\mcitedefaultmidpunct}
{\mcitedefaultendpunct}{\mcitedefaultseppunct}\relax
\EndOfBibitem
\bibitem[Tao \latin{et~al.}(2003)Tao, Perdew, Staroverov, and
  Scuseria]{tao2003climbing}
Tao,~J.; Perdew,~J.~P.; Staroverov,~V.~N.; Scuseria,~G.~E. Climbing the density
  functional ladder: Nonempirical meta--generalized gradient approximation
  designed for molecules and solids. \emph{Physical Review Letters}
  \textbf{2003}, \emph{91}, 146401\relax
\mciteBstWouldAddEndPuncttrue
\mciteSetBstMidEndSepPunct{\mcitedefaultmidpunct}
{\mcitedefaultendpunct}{\mcitedefaultseppunct}\relax
\EndOfBibitem
\bibitem[Sch{\"a}fer \latin{et~al.}(1992)Sch{\"a}fer, Horn, and
  Ahlrichs]{schafer1992fully}
Sch{\"a}fer,~A.; Horn,~H.; Ahlrichs,~R. Fully optimized contracted Gaussian
  basis sets for atoms Li to Kr. \emph{The Journal of Chemical Physics}
  \textbf{1992}, \emph{97}, 2571--2577\relax
\mciteBstWouldAddEndPuncttrue
\mciteSetBstMidEndSepPunct{\mcitedefaultmidpunct}
{\mcitedefaultendpunct}{\mcitedefaultseppunct}\relax
\EndOfBibitem
\bibitem[Grimme \latin{et~al.}(2010)Grimme, Antony, Ehrlich, and
  Krieg]{grimme2010consistent}
Grimme,~S.; Antony,~J.; Ehrlich,~S.; Krieg,~H. A consistent and accurate ab
  initio parametrization of density functional dispersion correction (DFT-D)
  for the 94 elements H-Pu. \emph{The Journal of chemical physics}
  \textbf{2010}, \emph{132}, 154104\relax
\mciteBstWouldAddEndPuncttrue
\mciteSetBstMidEndSepPunct{\mcitedefaultmidpunct}
{\mcitedefaultendpunct}{\mcitedefaultseppunct}\relax
\EndOfBibitem
\bibitem[Bartlett and Musia{\l}(2007)Bartlett, and
  Musia{\l}]{bartlett2007coupled}
Bartlett,~R.~J.; Musia{\l},~M. Coupled-cluster theory in quantum chemistry.
  \emph{Reviews of Modern Physics} \textbf{2007}, \emph{79}, 291\relax
\mciteBstWouldAddEndPuncttrue
\mciteSetBstMidEndSepPunct{\mcitedefaultmidpunct}
{\mcitedefaultendpunct}{\mcitedefaultseppunct}\relax
\EndOfBibitem
\bibitem[Shavitt and Bartlett(2009)Shavitt, and Bartlett]{shavitt2009many}
Shavitt,~I.; Bartlett,~R.~J. \emph{Many-body methods in chemistry and physics:
  MBPT and coupled-cluster theory}; Cambridge university press, 2009\relax
\mciteBstWouldAddEndPuncttrue
\mciteSetBstMidEndSepPunct{\mcitedefaultmidpunct}
{\mcitedefaultendpunct}{\mcitedefaultseppunct}\relax
\EndOfBibitem
\bibitem[Sun \latin{et~al.}(2018)Sun, Berkelbach, Blunt, Booth, Guo, Li, Liu,
  McClain, Sayfutyarova, Sharma, \latin{et~al.} others]{sun2018pyscf}
Sun,~Q.; Berkelbach,~T.~C.; Blunt,~N.~S.; Booth,~G.~H.; Guo,~S.; Li,~Z.;
  Liu,~J.; McClain,~J.~D.; Sayfutyarova,~E.~R.; Sharma,~S., \latin{et~al.}
  PySCF: the Python-based simulations of chemistry framework. \emph{Wiley
  Interdisciplinary Reviews: Computational Molecular Science} \textbf{2018},
  \emph{8}, e1340\relax
\mciteBstWouldAddEndPuncttrue
\mciteSetBstMidEndSepPunct{\mcitedefaultmidpunct}
{\mcitedefaultendpunct}{\mcitedefaultseppunct}\relax
\EndOfBibitem
\bibitem[Sun \latin{et~al.}(2020)Sun, Zhang, Banerjee, Bao, Barbry, Blunt,
  Bogdanov, Booth, Chen, Cui, \latin{et~al.} others]{sun2020recent}
Sun,~Q.; Zhang,~X.; Banerjee,~S.; Bao,~P.; Barbry,~M.; Blunt,~N.~S.;
  Bogdanov,~N.~A.; Booth,~G.~H.; Chen,~J.; Cui,~Z.-H., \latin{et~al.}  Recent
  developments in the PySCF program package. \emph{The Journal of chemical
  physics} \textbf{2020}, \emph{153}, 024109\relax
\mciteBstWouldAddEndPuncttrue
\mciteSetBstMidEndSepPunct{\mcitedefaultmidpunct}
{\mcitedefaultendpunct}{\mcitedefaultseppunct}\relax
\EndOfBibitem
\bibitem[Saue(2011)]{saue2011relativistic}
Saue,~T. Relativistic Hamiltonians for chemistry: A primer. \emph{ChemPhysChem}
  \textbf{2011}, \emph{12}, 3077--3094\relax
\mciteBstWouldAddEndPuncttrue
\mciteSetBstMidEndSepPunct{\mcitedefaultmidpunct}
{\mcitedefaultendpunct}{\mcitedefaultseppunct}\relax
\EndOfBibitem
\bibitem[Peng and Reiher(2012)Peng, and Reiher]{peng2012exact}
Peng,~D.; Reiher,~M. Exact decoupling of the relativistic Fock operator.
  \emph{Theoretical Chemistry Accounts} \textbf{2012}, \emph{131}, 1--20\relax
\mciteBstWouldAddEndPuncttrue
\mciteSetBstMidEndSepPunct{\mcitedefaultmidpunct}
{\mcitedefaultendpunct}{\mcitedefaultseppunct}\relax
\EndOfBibitem
\bibitem[Kutzelnigg and Liu(2005)Kutzelnigg, and
  Liu]{kutzelnigg2005quasirelativistic}
Kutzelnigg,~W.; Liu,~W. Quasirelativistic theory equivalent to fully
  relativistic theory. \emph{The Journal of chemical physics} \textbf{2005},
  \emph{123}, 241102\relax
\mciteBstWouldAddEndPuncttrue
\mciteSetBstMidEndSepPunct{\mcitedefaultmidpunct}
{\mcitedefaultendpunct}{\mcitedefaultseppunct}\relax
\EndOfBibitem
\bibitem[Becke(1988)]{becke1988density}
Becke,~A.~D. Density-functional exchange-energy approximation with correct
  asymptotic behavior. \emph{Physical review A} \textbf{1988}, \emph{38},
  3098\relax
\mciteBstWouldAddEndPuncttrue
\mciteSetBstMidEndSepPunct{\mcitedefaultmidpunct}
{\mcitedefaultendpunct}{\mcitedefaultseppunct}\relax
\EndOfBibitem
\bibitem[Lee \latin{et~al.}(1988)Lee, Yang, and Parr]{lee1988development}
Lee,~C.; Yang,~W.; Parr,~R.~G. Development of the Colle-Salvetti
  correlation-energy formula into a functional of the electron density.
  \emph{Physical review B} \textbf{1988}, \emph{37}, 785\relax
\mciteBstWouldAddEndPuncttrue
\mciteSetBstMidEndSepPunct{\mcitedefaultmidpunct}
{\mcitedefaultendpunct}{\mcitedefaultseppunct}\relax
\EndOfBibitem
\bibitem[Becke(1993)]{becke1993new}
Becke,~A.~D. A new mixing of Hartree--Fock and local density-functional
  theories. \emph{The Journal of chemical physics} \textbf{1993}, \emph{98},
  1372--1377\relax
\mciteBstWouldAddEndPuncttrue
\mciteSetBstMidEndSepPunct{\mcitedefaultmidpunct}
{\mcitedefaultendpunct}{\mcitedefaultseppunct}\relax
\EndOfBibitem
\bibitem[Schurkus \latin{et~al.}(2020)Schurkus, Chen, Cheng, Chan, and
  Stanton]{schurkus2020theoretical}
Schurkus,~H.; Chen,~D.-T.; Cheng,~H.-P.; Chan,~G.; Stanton,~J. Theoretical
  prediction of magnetic exchange coupling constants from broken-symmetry
  coupled cluster calculations. \emph{The Journal of Chemical Physics}
  \textbf{2020}, \emph{152}, 234115\relax
\mciteBstWouldAddEndPuncttrue
\mciteSetBstMidEndSepPunct{\mcitedefaultmidpunct}
{\mcitedefaultendpunct}{\mcitedefaultseppunct}\relax
\EndOfBibitem
\bibitem[Yamaguchi \latin{et~al.}(1986)Yamaguchi, Takahara, and
  Fueno]{yamaguchi1986ab}
Yamaguchi,~K.; Takahara,~Y.; Fueno,~T. Ab-initio molecular orbital studies of
  structure and reactivity of transition metal-oxo compounds. Applied Quantum
  Chemistry: Proceedings of the Nobel Laureate Symposium on Applied Quantum
  Chemistry in Honor of G. Herzberg, RS Mulliken, K. Fukui, W. Lipscomb, and R.
  Hoffman, Honolulu, HI, 16--21 December 1984. 1986; pp 155--184\relax
\mciteBstWouldAddEndPuncttrue
\mciteSetBstMidEndSepPunct{\mcitedefaultmidpunct}
{\mcitedefaultendpunct}{\mcitedefaultseppunct}\relax
\EndOfBibitem
\bibitem[Pipek and Mezey(1989)Pipek, and Mezey]{pipek1989fast}
Pipek,~J.; Mezey,~P.~G. A fast intrinsic localization procedure applicable for
  abinitio and semiempirical linear combination of atomic orbital wave
  functions. \emph{The Journal of Chemical Physics} \textbf{1989}, \emph{90},
  4916--4926\relax
\mciteBstWouldAddEndPuncttrue
\mciteSetBstMidEndSepPunct{\mcitedefaultmidpunct}
{\mcitedefaultendpunct}{\mcitedefaultseppunct}\relax
\EndOfBibitem
\bibitem[Lee \latin{et~al.}(2021)Lee, Zhai, Sharma, Umrigar, and
  Chan]{lee2021externally}
Lee,~S.; Zhai,~H.; Sharma,~S.; Umrigar,~C.~J.; Chan,~G. K.-L. Externally
  Corrected CCSD with Renormalized Perturbative Triples (R-ecCCSD (T)) and the
  Density Matrix Renormalization Group and Selected Configuration Interaction
  External Sources. \emph{Journal of Chemical Theory and Computation}
  \textbf{2021}, \emph{17}, 3414--3425\relax
\mciteBstWouldAddEndPuncttrue
\mciteSetBstMidEndSepPunct{\mcitedefaultmidpunct}
{\mcitedefaultendpunct}{\mcitedefaultseppunct}\relax
\EndOfBibitem
\bibitem[M{\o}ller and Plesset(1934)M{\o}ller, and Plesset]{moller1934note}
M{\o}ller,~C.; Plesset,~M.~S. Note on an approximation treatment for
  many-electron systems. \emph{Physical review} \textbf{1934}, \emph{46},
  618\relax
\mciteBstWouldAddEndPuncttrue
\mciteSetBstMidEndSepPunct{\mcitedefaultmidpunct}
{\mcitedefaultendpunct}{\mcitedefaultseppunct}\relax
\EndOfBibitem
\bibitem[Pople \latin{et~al.}(1976)Pople, Binkley, and
  Seeger]{pople1976theoretical}
Pople,~J.~A.; Binkley,~J.~S.; Seeger,~R. Theoretical models incorporating
  electron correlation. \emph{International Journal of Quantum Chemistry}
  \textbf{1976}, \emph{10}, 1--19\relax
\mciteBstWouldAddEndPuncttrue
\mciteSetBstMidEndSepPunct{\mcitedefaultmidpunct}
{\mcitedefaultendpunct}{\mcitedefaultseppunct}\relax
\EndOfBibitem
\bibitem[Neese and Valeev(2011)Neese, and Valeev]{neese2011revisiting}
Neese,~F.; Valeev,~E.~F. Revisiting the atomic natural orbital approach for
  basis sets: robust systematic basis sets for explicitly correlated and
  conventional correlated ab initio methods? \emph{Journal of chemical theory
  and computation} \textbf{2011}, \emph{7}, 33--43\relax
\mciteBstWouldAddEndPuncttrue
\mciteSetBstMidEndSepPunct{\mcitedefaultmidpunct}
{\mcitedefaultendpunct}{\mcitedefaultseppunct}\relax
\EndOfBibitem
\bibitem[Perdew \latin{et~al.}(1996)Perdew, Burke, and
  Ernzerhof]{perdew1996generalized}
Perdew,~J.~P.; Burke,~K.; Ernzerhof,~M. Generalized gradient approximation made
  simple. \emph{Physical review letters} \textbf{1996}, \emph{77}, 3865\relax
\mciteBstWouldAddEndPuncttrue
\mciteSetBstMidEndSepPunct{\mcitedefaultmidpunct}
{\mcitedefaultendpunct}{\mcitedefaultseppunct}\relax
\EndOfBibitem
\bibitem[Grimme(2006)]{grimme2006semiempirical}
Grimme,~S. Semiempirical GGA-type density functional constructed with a
  long-range dispersion correction. \emph{Journal of computational chemistry}
  \textbf{2006}, \emph{27}, 1787--1799\relax
\mciteBstWouldAddEndPuncttrue
\mciteSetBstMidEndSepPunct{\mcitedefaultmidpunct}
{\mcitedefaultendpunct}{\mcitedefaultseppunct}\relax
\EndOfBibitem
\bibitem[Furness \latin{et~al.}(2020)Furness, Kaplan, Ning, Perdew, and
  Sun]{furness2020accurate}
Furness,~J.~W.; Kaplan,~A.~D.; Ning,~J.; Perdew,~J.~P.; Sun,~J. Accurate and
  numerically efficient r2SCAN meta-generalized gradient approximation.
  \emph{The journal of physical chemistry letters} \textbf{2020}, \emph{11},
  8208--8215\relax
\mciteBstWouldAddEndPuncttrue
\mciteSetBstMidEndSepPunct{\mcitedefaultmidpunct}
{\mcitedefaultendpunct}{\mcitedefaultseppunct}\relax
\EndOfBibitem
\bibitem[Staroverov \latin{et~al.}(2003)Staroverov, Scuseria, Tao, and
  Perdew]{staroverov2003comparative}
Staroverov,~V.~N.; Scuseria,~G.~E.; Tao,~J.; Perdew,~J.~P. Comparative
  assessment of a new nonempirical density functional: Molecules and
  hydrogen-bonded complexes. \emph{The Journal of chemical physics}
  \textbf{2003}, \emph{119}, 12129--12137\relax
\mciteBstWouldAddEndPuncttrue
\mciteSetBstMidEndSepPunct{\mcitedefaultmidpunct}
{\mcitedefaultendpunct}{\mcitedefaultseppunct}\relax
\EndOfBibitem
\bibitem[Salomon \latin{et~al.}(2002)Salomon, Reiher, and
  Hess]{salomon2002assertion}
Salomon,~O.; Reiher,~M.; Hess,~B.~A. Assertion and validation of the
  performance of the B3LYP* functional for the first transition metal row and
  the G2 test set. \emph{The Journal of chemical physics} \textbf{2002},
  \emph{117}, 4729--4737\relax
\mciteBstWouldAddEndPuncttrue
\mciteSetBstMidEndSepPunct{\mcitedefaultmidpunct}
{\mcitedefaultendpunct}{\mcitedefaultseppunct}\relax
\EndOfBibitem
\bibitem[Perdew \latin{et~al.}(1996)Perdew, Ernzerhof, and
  Burke]{perdew1996rationale}
Perdew,~J.~P.; Ernzerhof,~M.; Burke,~K. Rationale for mixing exact exchange
  with density functional approximations. \emph{The Journal of chemical
  physics} \textbf{1996}, \emph{105}, 9982--9985\relax
\mciteBstWouldAddEndPuncttrue
\mciteSetBstMidEndSepPunct{\mcitedefaultmidpunct}
{\mcitedefaultendpunct}{\mcitedefaultseppunct}\relax
\EndOfBibitem
\bibitem[Zhao and Truhlar(2008)Zhao, and Truhlar]{zhao2008m06}
Zhao,~Y.; Truhlar,~D.~G. The M06 suite of density functionals for main group
  thermochemistry, thermochemical kinetics, noncovalent interactions, excited
  states, and transition elements: two new functionals and systematic testing
  of four M06-class functionals and 12 other functionals. \emph{Theoretical
  chemistry accounts} \textbf{2008}, \emph{120}, 215--241\relax
\mciteBstWouldAddEndPuncttrue
\mciteSetBstMidEndSepPunct{\mcitedefaultmidpunct}
{\mcitedefaultendpunct}{\mcitedefaultseppunct}\relax
\EndOfBibitem
\bibitem[Canc{\`e}s \latin{et~al.}(2013)Canc{\`e}s, Maday, and
  Stamm]{cances2013domain}
Canc{\`e}s,~E.; Maday,~Y.; Stamm,~B. Domain decomposition for implicit
  solvation models. \emph{The Journal of Chemical Physics} \textbf{2013},
  \emph{139}, 054111\relax
\mciteBstWouldAddEndPuncttrue
\mciteSetBstMidEndSepPunct{\mcitedefaultmidpunct}
{\mcitedefaultendpunct}{\mcitedefaultseppunct}\relax
\EndOfBibitem
\end{mcitethebibliography}

\onecolumngrid

\section*{TOC Graphic}
\includegraphics[width=400px]{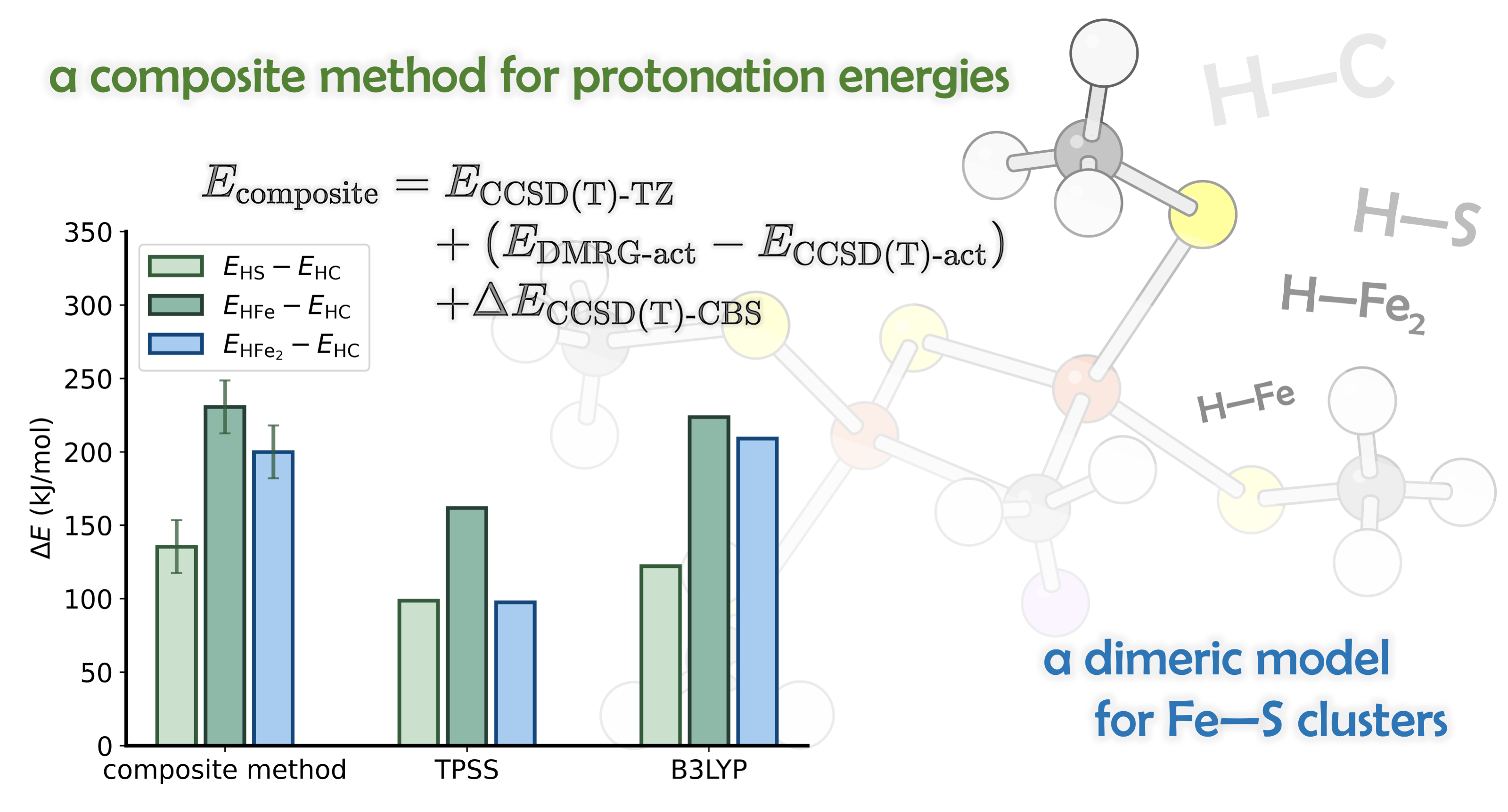}

\end{document}